\documentclass[reprint,superscriptaddress,amsmath,amssymb, aps,prb,floatfix]{revtex4-2}
\usepackage{graphicx}
\usepackage{dcolumn}
\usepackage{bm}
\begin{document}

\title{Thermally-controlled flux avalanche dynamics in bulk NbTi superconductor}
\author{I. Abaloszewa}
\email{abali@ifpan.edu.pl}
\affiliation{Institute of Physics, Polish Academy of Sciences, Warsaw, Poland}
\author{V. V. Chabanenko}
\affiliation{O. Galkin Donetsk Institute for Physics and Engineering, National Academy of Science, Kyiv, Ukraine}
\author{A. Abaloszew}
\affiliation{Institute of Physics, Polish Academy of Sciences, Warsaw, Poland}
\date{\today}

\begin{abstract}
We report the first direct visualization of flux avalanche propagation dynamics in bulk superconducting NbTi, tracking individual events and measuring their velocities using high-speed magneto-optical imaging. Unlike thin films with electromagnetic avalanches at km/s speeds, we observe velocities of 15--25 m/s, which are orders of magnitude slower. Analysis of characteristic timescales reveals that these avalanches are governed by local heating and limited heat dissipation through the adhesive layer, establishing a fundamentally different, thermally limited propagation regime. The threshold field for avalanche nucleation decreases with temperature, contrary to the increasing trend in thin films with efficient cooling - a behavior consistent with slow heat removal and thermal runaway in our system. All observed avalanches exhibit universal normalized velocity-distance scaling despite varying morphologies, confirming the robustness of thermal control. These findings reveal that bulk superconductors with poor thermal coupling operate in a previously uncharacterized avalanche regime, with direct implications for flux stability and quench protection in NbTi-based magnets, as well as a broader understanding of thermomagnetic instabilities in technological superconductors.
\end{abstract}

\maketitle

\section{Introduction}

For decades, intensive studies have focused on thermomagnetic instabilities, which manifest as sudden avalanches of magnetic flux in superconducting materials, both in thin-film and bulk superconductors. In a type II superconductor, when the external magnetic field exceeds the level equal to the lower critical field $H_{\text{c1}}$, the magnetic flux enters the superconductor in the form of quantized flux lines, i.e., vortices. As the external field increases, the vortices gradually penetrate inside the superconductor, causing a local heat release due to the motion of the normal vortex core. In real superconductors, vortices interact with defects (pinning centers), preventing motion and creating non-uniform flux. Local fluctuations can then trigger a positive feedback process, where insufficient heat dissipation leads to local overheating. This overheating reduces the pinning force of the vortices in these areas or even causes a local transition of the superconductor to a normal state, promoting rapid movement of the magnetic flux into the sample in the form of avalanches. So far, such thermomagnetic instabilities have been observed in numerous superconducting materials, such as, for example, YBa$_2$Cu$_3$O$_{7-\delta}$ \cite{Baziljevich}, MgB$_2$ \cite{Johansen_2002}, Pb \cite{Awad}, Nb \cite{Wertheimer}, Nb$_3$Sn \cite{Rudnev1}, NbN \cite{Rudnev2}, NbTi \cite{Vasiliev}, NbTiN \cite{Nulens}, a-MoGe \cite{Motta}, or YNi$_2$B$_2$C \cite{Wimbush}. While thin-film superconductors have been extensively studied using magneto-optical imaging (MOI), direct time-resolved measurements of flux avalanches in bulk materials remain scarce. In particular, NbTi - the most widely used material for superconducting magnets - has not been investigated dynamically by MOI, despite its importance for flux stability and quench protection.

A key issue in avalanche dynamics is determining their propagation velocity. This parameter serves as a sensitive indicator of the dominant physical mechanism of instability and has practical significance, for example, in estimating the critical response times of quench protection systems in superconducting devices. However, experimental observations on the temporal evolution of the avalanche process remain extremely rare. Previously, a spatio-temporal study was carried out to measure the avalanche velocity in a 20~$\mu$m thick Nb foil by comparing the signals from Hall sensors placed on the foil 50~$\mu$m apart \cite{Behnia}. The average avalanche transit time was estimated to be 0.8~ms, which gives an avalanche velocity of a few~cm/s. In \cite{Harrison}, flux jumps were studied in bulk samples of Nb and NbZr using magneto-optics and high-speed cinematography, yielding flux front velocities in the initial propagation phase of approximately 5~m/s and 7.5~m/s, respectively. In other materials, such as bulk NbTi, MgB$_2$ or Bi$_2$Sr$_2$CaCu$_2$O$_{8+\delta}$, the avalanche velocities measured using inductive and Hall sensors were 11~m/s, 16~m/s, 70~m/s respectively \cite{Ch1, Ch2, Ch3}. Interestingly, in thin films of MgB$_2$ or YBa$_2$Cu$_3$O$_{7-\delta}$, the dendrite avalanche propagation velocities were estimated to be approximately 14--25~km/s, and there, the study of avalanche development over time represents a non-trivial task, which was accomplished using a femtosecond pulsed laser as a trigger for synchronizing the recording of images of the flux distribution during the avalanche propagation \cite{Vestgarden, Bolz, Bolz2}. These large values exceed the speed of sound in these materials and are orders of magnitude higher than the avalanche propagation velocities in niobium foil and bulk materials reported in \cite{Behnia, Harrison, Ch1, Ch2, Ch3}. Such discrepancies point to different scenarios of thermomagnetic instability propagation. It is also useful to know the temperature dependence of the threshold field for the first instability $H_{\text{th}}$, because it determines the ``window'' for the application of materials where no instabilities are observed. 

In this work, we present the first direct observation of the spatiotemporal evolution of magnetic flux avalanches in a NbTi disk. Moreover, we have examined for the first time the temperature dependence of the thermomagnetic threshold field $H_{\text{th}}(T)$ for this type of specimen, which we compare to the corresponding dependence in thin-film samples. While performing these experiments, we encountered a limitation in the applied magnetic field ramp rate at low temperatures in our measurement system. To properly analyze dynamic processes, it is crucial to distinguish between timescales intrinsic to the material and those imposed by experimental conditions; therefore, we characterized how our measurement system affects the magnetic field ramp rate at the sample location throughout the relevant temperature range.

\section{Sample and experimental setup}

The disk-shaped sample of NbTi 50 at$\%$ alloy, with a thickness of $d = 0.1$~mm and a diameter of 12~mm (radius $R = 6$~mm) was prepared according to the procedure described in \cite{Ch4}.

The research was carried out using the MOI method. This method for studying superconducting materials is based on imaging the magnetic flux penetration into a superconducting sample cooled below the superconducting critical temperature. The sample is mounted on a sample holder connected to a liquid helium-cooled cold finger in an optical cryostat chamber pumped to high vacuum. The sample is attached to the cold finger of the sample holder using nonadecane (C$_{19}$H$_{40}$). It solidifies at $\sim$305~K and provides mechanical adhesion and thermal contact that is better than commonly used cryogenic greases such as Apiezon \cite{nonadecane, ApiezonN}). A pair of Helmholtz coils, mounted outside the optical cryostat, is used to apply a uniform external magnetic field perpendicular to the sample plane during the measurements.

The magnetic flux in the superconducting sample is visualized using the Faraday effect in a Bi-doped ferrite garnet indicator film with in-plane magnetization, placed on top of the sample. The indicator rotates the plane of polarized light formed by the polarizer, and the angle of rotation is proportional to the local magnetic field parallel to the direction of light propagation. The image of polarized light intensity corresponding to the magnetic flux distribution in the sample is recorded by a setup consisting of a polarization microscope, digital camera, and computer.

In our experiment, we used a high-resolution Hamamatsu ORCA II ERG camera and a high-speed Phantom VEO 710 camera capable of recording at a maximum frame rate of 22,000~fps. Magnetic field measurements at the sample location on the cold finger were performed with a Toshiba THS118 GaAs Hall sensor.

\begin{figure}
\includegraphics[width=8.5cm]{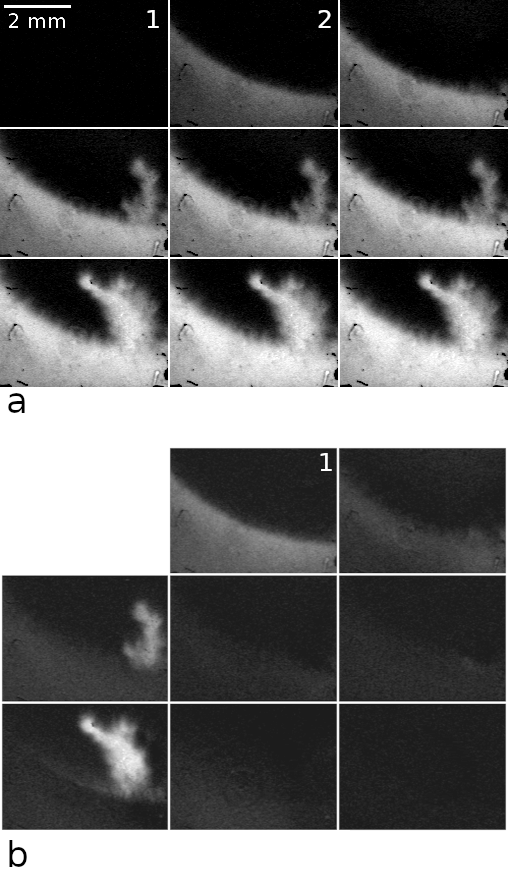}
 \caption{(a) Magneto-optical visualization of the NbTi disk edge at 5 K. An external magnetic field of 50~mT is switched on between the first and second frames. Frames are captured every 45~ms. (b) Each image shows the difference between neighboring frames in (a), for example, the image marked with the number~1 in (b) represents the difference between frames 2 and 1 in (a).}
 \label{Fig1}
\end{figure}

\section{Results}

\subsection{High-resolution camera: temporal resolution and limitations}

During the application of an external magnetic field to our sample-indicator system, we deal with several parallel processes that contribute to the overall dynamic picture visualized by magneto-optics. Firstly, the external magnetic field ramps up at a finite rate. Secondly, the sample responds at its own characteristic timescale, involving magnetic flux propagation in the form of a smooth front or abrupt avalanches, the establishment of shielding currents, and the generation of a demagnetizing field. Finally, the indicator reacts to changes in the local magnetic field by rotating the plane of polarization at a finite rate.

To measure the actual field sweep rate at the sample location and obtain images of the magnetic flux penetration into the sample as a function of time, we monitored field establishment using a Hamamatsu camera after applying a magnetic field via the magnet coil. The results are presented in Figure \ref{Fig1}. In this figure, we have visualized the penetration of magnetic flux into the NbTi disk at a temperature of 5~K by applying an external magnetic field $H_{\text{a}}$=50~mT.

The magneto-optical image shows part of the sample and the surrounding area. To visualize only the brightness changes caused by the applied magnetic field, a reference image obtained at zero external field was subtracted from all images. This procedure excludes the effects of indicator defects and illumination inhomogeneity. The brightness outside the sample increases with the growing external field, reflecting field penetration into the surrounding space. The central region of the sample maintains the Meissner state with complete magnetic flux exclusion and appears as the darkest area. The edge regions of the sample appear brighter, indicating magnetic flux penetration.

It can be seen that the magnetic flux penetrates into the sample both as a relatively smooth front and as irregularly shaped avalanches. We find that the avalanches develop faster than the inter-frame interval between two frames recorded at the highest speed that our camera can provide, namely 45~ms. At this recording speed of 22~frames per second (fps), the image resolution is significantly reduced as the frame size of the camera is downsized to only $168\times128$ pixels. We do not observe the gradual development of individual avalanches; however, the relaxation processes accompanying the application of an external magnetic field span approximately nine such frames, i.e., about 0.4~s. During this time, avalanches appear from frame to frame with the proceeding flux entry. This is better visualized in Figure \ref{Fig1}(b), in which the images show the differences between two neighboring images of Figure \ref{Fig1}(a). We see that new avalanches appear as bright spots in the subsequent frames, but the existing avalanches do not change their length, i.e., the avalanches fully develop in a time shorter than 45~ms.

\subsection{System-imposed field ramp rate limitations}
\label{Ramp}

\begin{figure}
\includegraphics[width=8.5cm]{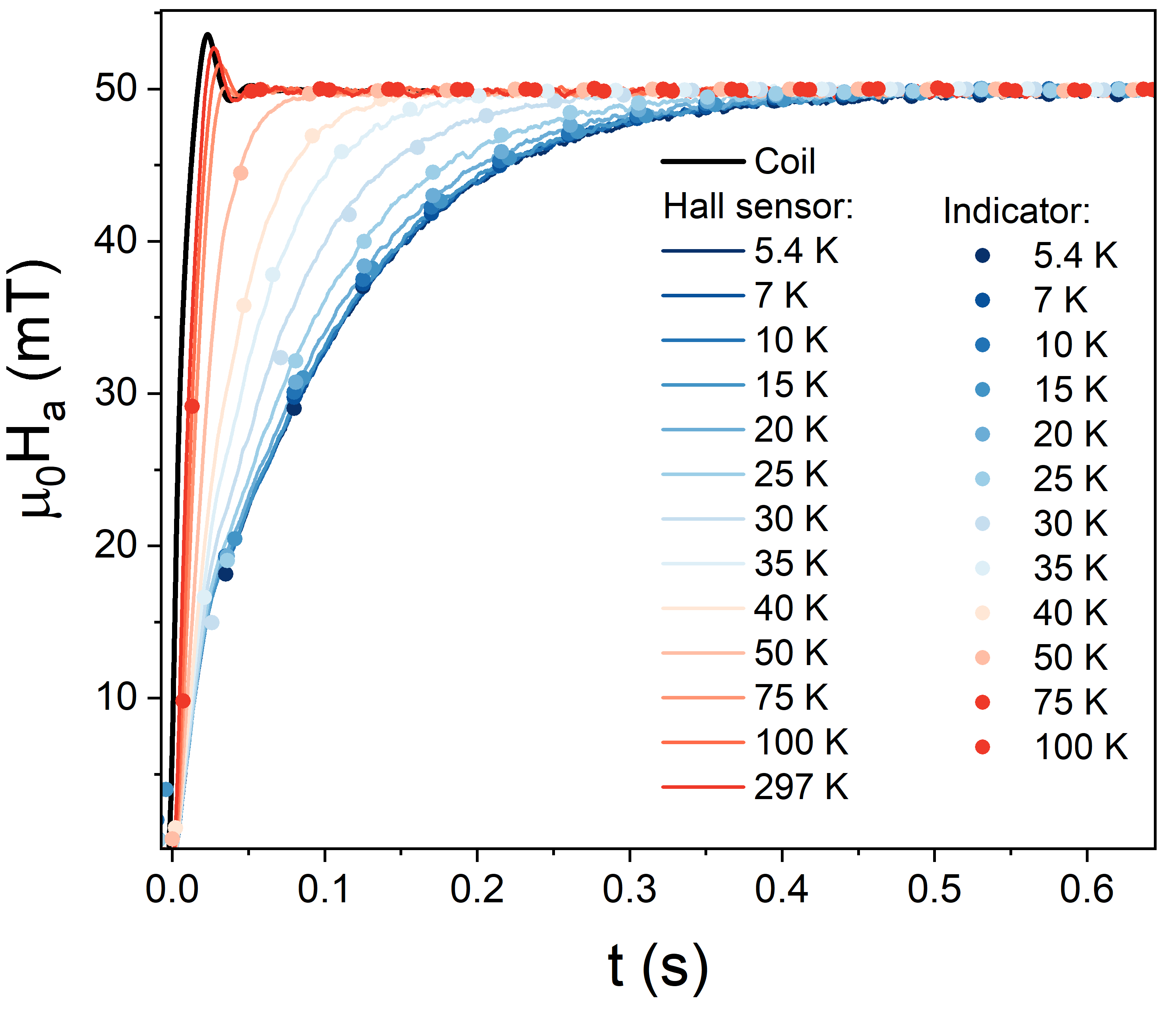}
\caption{Time evolution of the local magnetic field at the sample holder inside the cryostat during ramp-up of the external field generated by the coils. Measurements by indicator (symbols) and Hall sensor (lines) at various temperatures.}
\label{Fig2}
\end{figure}

At room temperature, the nominal magnetic field generated by our Helmholtz coils rises to 50~mT in approximately 0.04~s, corresponding to a sweep rate of $\sim$1.25~T/s. However, Figure \ref{Fig1} demonstrates that upon such field application, the field around the sample at 5~K establishes over a substantially longer timescale. Therefore, to understand whether this change is influenced by processes occurring in the sample or is exclusively due to eddy currents in the metallic parts of the cryostat, we measured the magnetic field rise at the sample position on the cold finger as a function of temperature. Measurements were performed over a temperature range of 5--300~K. Figure \ref{Fig2} compares the magnetic field rise at different temperatures at the sample location measured by MOI and by the Hall sensor, and in the coil at room temperature.

At room temperature, the field rise recorded by the Hall sensor and MOI indicator coincides with the oscilloscope readings showing the current increase in the coil and takes about 0.04~s. When cooled to a temperature of 5~K, the field establishment time increases significantly to about 0.4~s. The maximum field sweep rate, which occurs at the beginning of the ramp, is approximately 0.25~T/s at low temperatures. This temperature-dependent slowdown is attributed to eddy currents induced in the metallic components of the cryostat. As temperature decreases, the electrical resistance of these metal components drops significantly (with residual resistance ratios reaching 100--1000 for cryostat materials such as copper and aluminum \cite{Calatroni}), leading to stronger and longer-lasting screening currents that impede magnetic field penetration into the cryostat interior.

The Hall sensor and MOI readings agree perfectly. This agreement confirms that both measurement techniques retain sufficiently fast response times at cryogenic temperatures compared to the 0.1--0.4~s field establishment timescale observed in our experiments: GaAs Hall sensors, with their high electron mobility of $\sim$8500~cm$^2$/(V$\cdot$s), maintain reliable operation and fast response at temperatures down to 1.5~K \cite{Sawada}, while iron garnet magneto-optical films demonstrate high-speed response with nanosecond-scale \cite{Bolz, Bolz2}.

This temperature-dependent slowdown in the magnetic field evolution due to eddy currents in the cryostat components must be taken into account when studying dynamic phenomena in materials at low temperatures. The actual field sweep rate at the sample location should be determined at each temperature of interest rather than assuming it matches the nominal rate at the magnet coil. Failure to account for this effect can lead to incorrect conclusions about the intrinsic dynamics of the material under study. For example, the observed rate of magnetic flux penetration into a superconductor may be misinterpreted, with the delay introduced by the electromagnetic screening in the cryostat being mistaken for phenomena inherent to the superconducting material itself. Excellent work on minimizing the influence of parasitic eddy currents in the metallic parts of a cryostat surrounding the sample was reported by Baziljevich et al. \cite{Baziljevich2}.
In our study, the field-ramp rates were relatively low, and no special measures to suppress such effects were required. However, in situations where an estimation of the rate of change of the magnetic field applied to the sample was required, possible delays or distortions due to induced currents in the cryostat were taken into account.

\subsection{High-speed camera measurements}

\begin{figure}
\includegraphics[width=8.5cm]{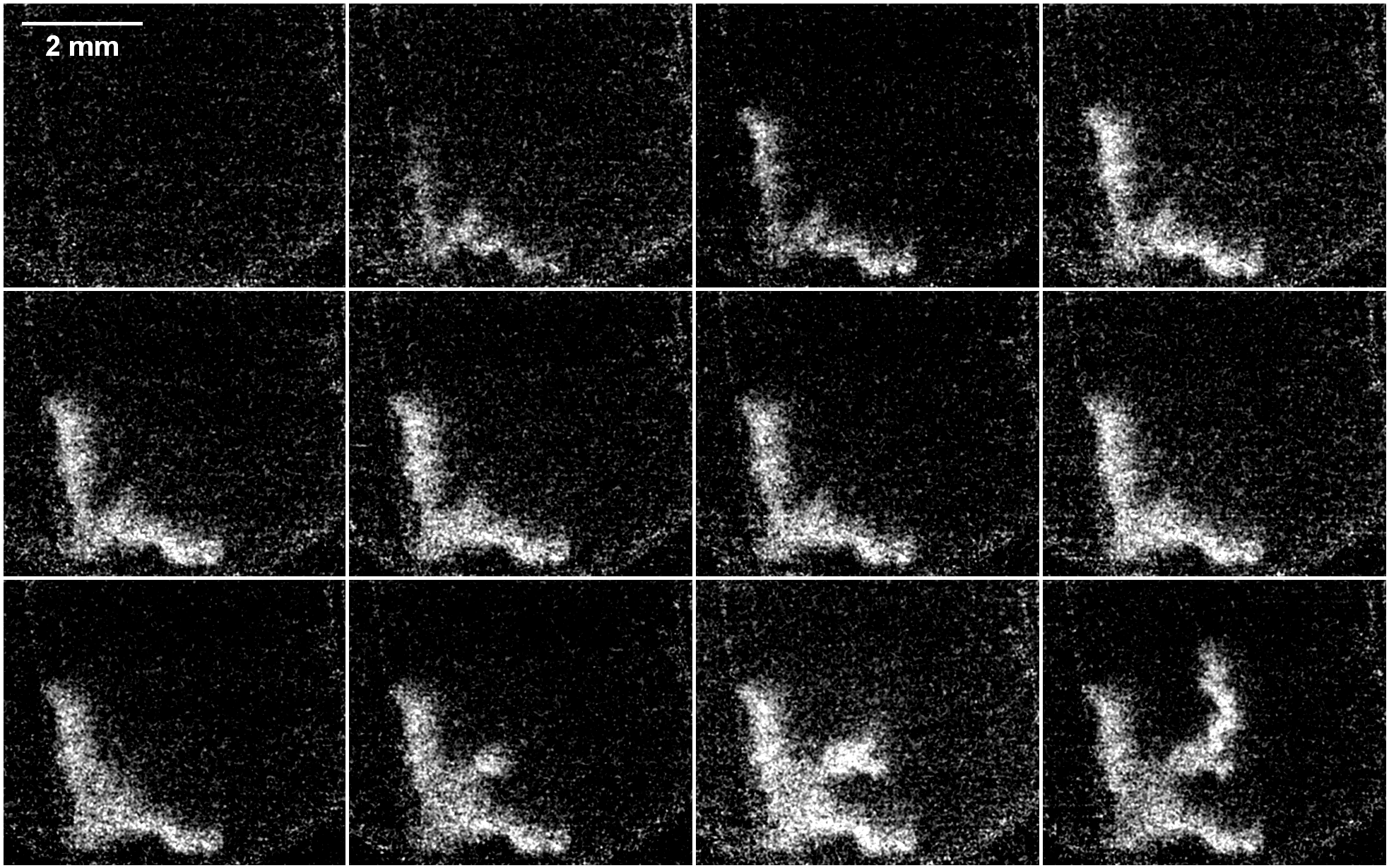}
\caption{The process of thermomagnetic avalanche entry recorded by a Phantom fast camera: 11,000 fps.}
\label{Fig3}
\end{figure}

To temporally resolve the avalanche-like penetration of magnetic flux, we recorded the evolution of such thermomagnetic instabilities using a high-speed Phantom VEO camera.

\begin{figure}
\includegraphics[width=8.5cm]{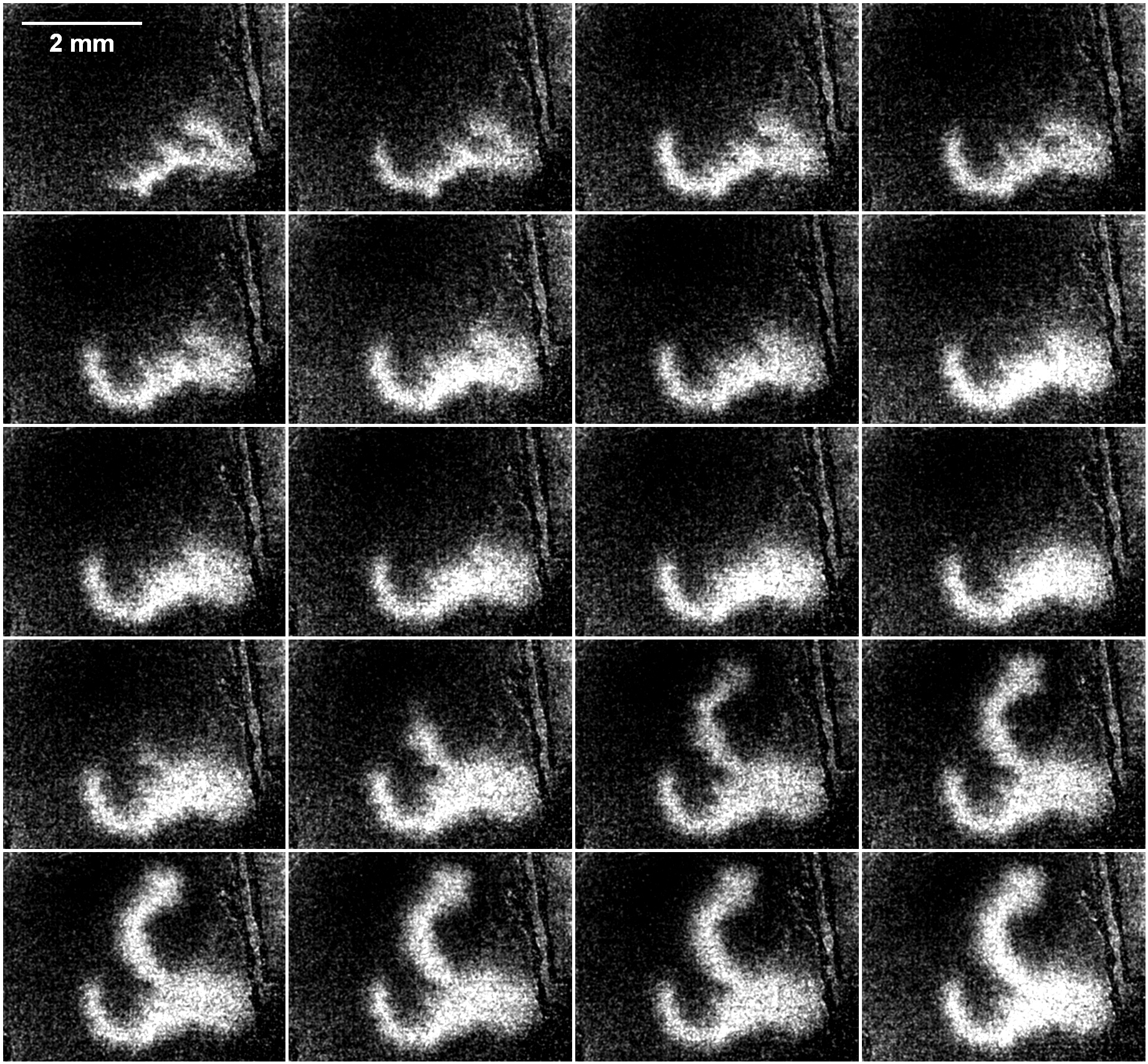}
\caption{The process of thermomagnetic avalanche entry recorded by a Phantom fast camera: 14,000~fps, first measurement.}
\label{Fig4}
\end{figure}

\begin{figure}
\includegraphics[width=8.5cm]{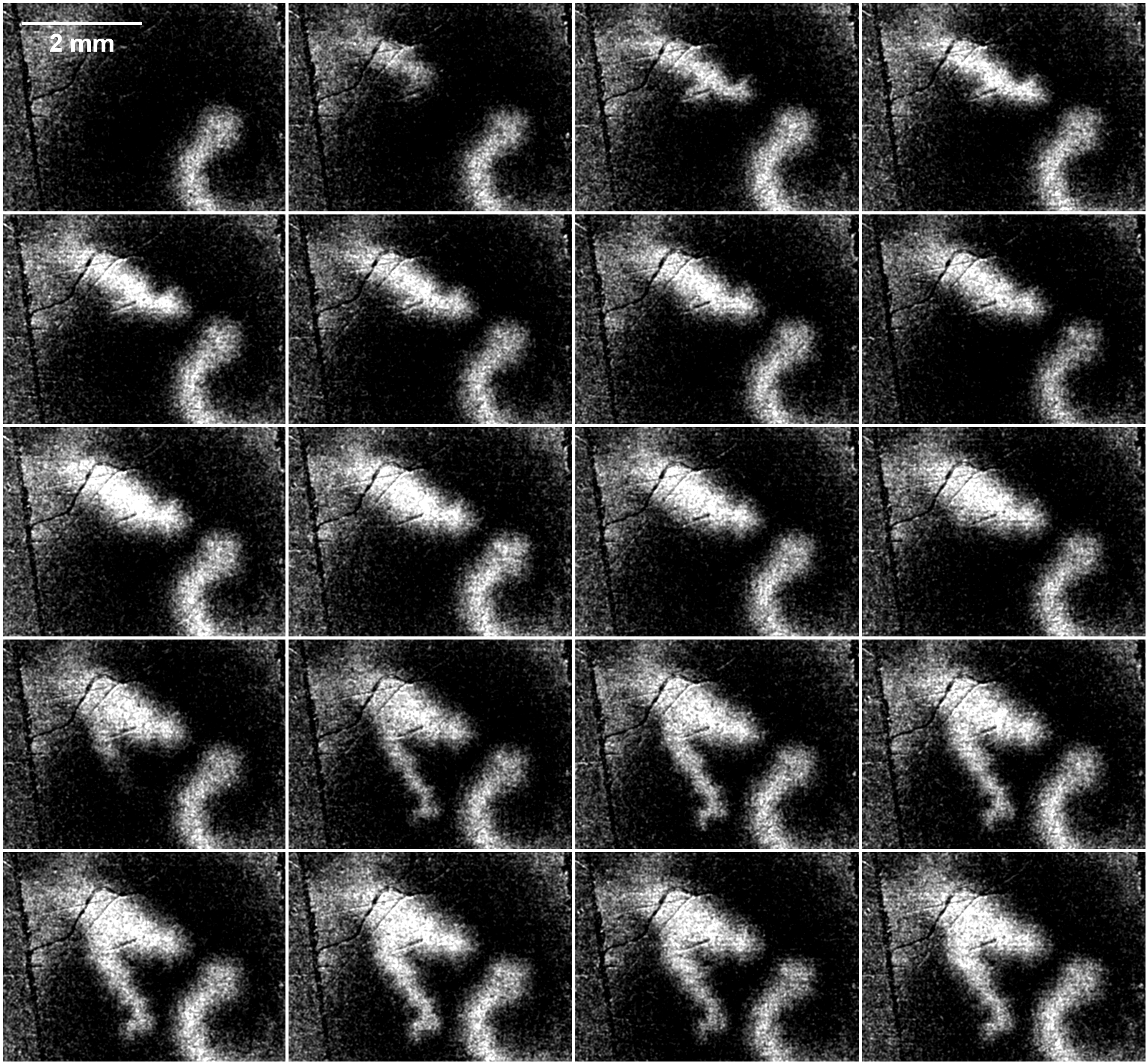}
\caption{The process of thermomagnetic avalanche entry recorded by a Phantom fast camera: 14,000~fps, second measurement.}
\label{Fig5}
\end{figure}

\begin{figure}
\includegraphics[width=8.5cm]{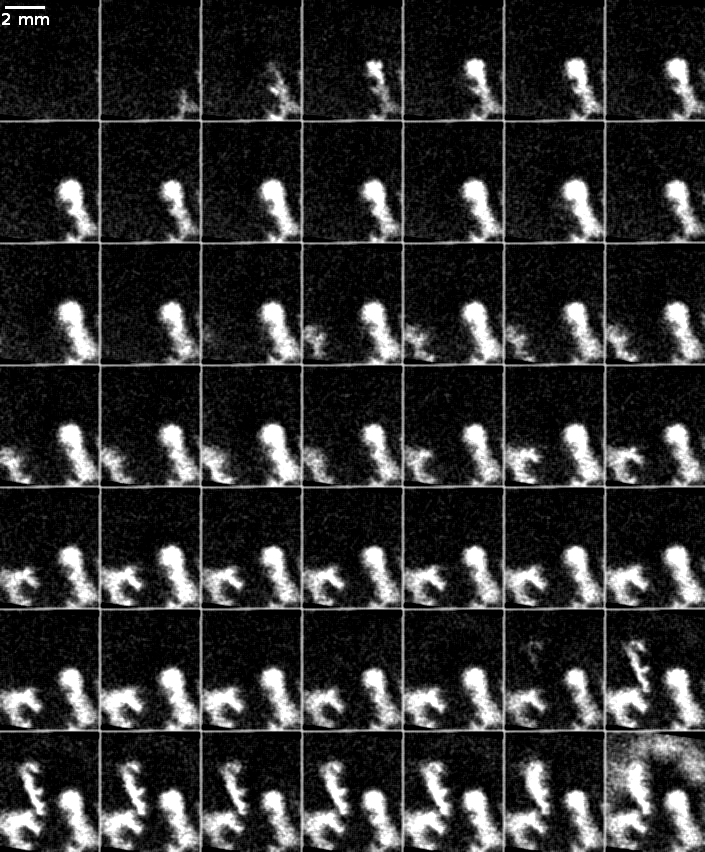}
\caption{The process of thermomagnetic avalanche entry recorded by a Phantom fast camera: 22,000~fps.}
\label{Fig6}
\end{figure}

Figures \ref{Fig3}-\ref{Fig6} show the evolution of avalanches in the disk at 6~K during ramping of the external magnetic field from 0 to 60~mT. Images were recorded at frame rates of 11,000, 14,000, and 22,000~fps, with the highest rate (Figure~\ref{Fig6}) providing superior temporal resolution for capturing the initial propagation phase. To enhance contrast and visualize the intensity changes caused by avalanche penetration, we subtracted the frame captured immediately before avalanche onset from subsequent frames.

The avalanche penetration process typically spans several consecutive frames, enabling quantitative analysis of the avalanche entry dynamics and propagation velocity. Between individual avalanche events, the recorded sequences contain tens to hundreds of frames without observable changes. Therefore, Figures \ref{Fig3}-\ref{Fig6} show only frames with visible flux penetration, omitting the frame sequences containing no additional information.

\begin{figure}
\includegraphics[width=8.5cm]{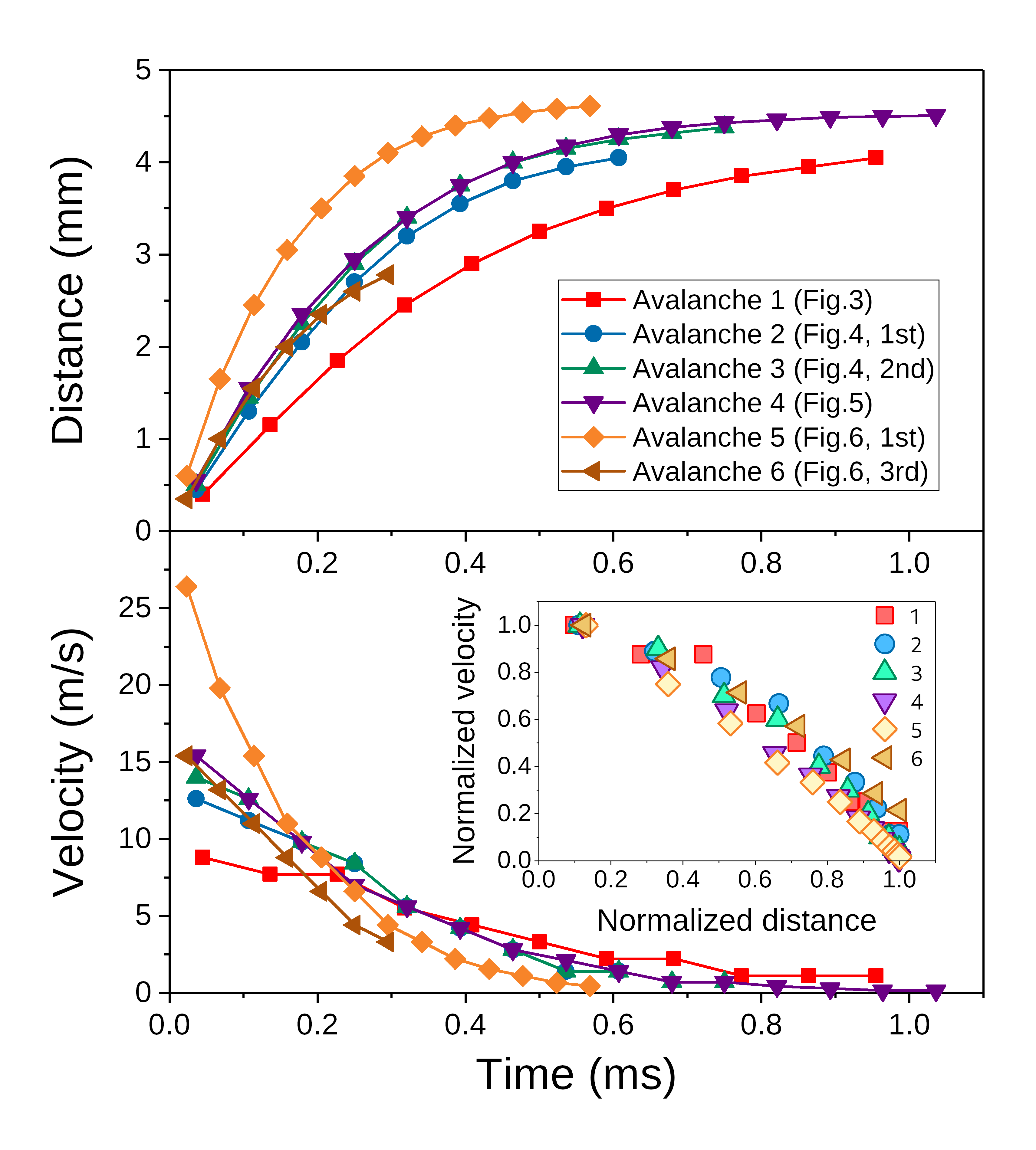}
\caption{Spatiotemporal analysis of flux avalanche propagation in 
the NbTi disk at 6~K. Upper panel: Penetration distance versus time for six avalanche events (numbered 1-6), selected from Figures~\ref{Fig3}-\ref{Fig6} for their clear visibility and representative propagation dynamics. Lower panel: Instantaneous velocity showing deceleration from 15--25~m/s to 5--10~m/s. Inset: Normalized velocity versus normalized distance, demonstrating universal scaling behavior independent of individual avalanche characteristics.}
\label{Fig7}
\end{figure}

Figure~\ref{Fig6} illustrates this approach: from an original sequence spanning approximately 25~ms, only 49 frames showing avalanche development are displayed (frame numbers: 5--18, 234--258, 535--542, 751, and 2333). At this frame rate, the time interval between consecutive frames is 45.44~$\mu$s. To quantitatively analyze avalanche dynamics, we tracked the spatial evolution of the avalanche front in six representative events captured in Figures~\ref{Fig3}-\ref{Fig6}. For each avalanche, we measured the penetration distance as a function of time, allowing us to extract both the time-dependent position and instantaneous velocity of the advancing flux front.

Figure~\ref{Fig7} presents a detailed analysis of avalanche propagation. The upper panel shows the distance traveled by the avalanche front - defined as the location of maximum brightness gradient - as a function of time for six different events. All avalanches exhibit similar behavior: rapid initial penetration followed by gradual deceleration as the front advances into the sample. The fronts typically penetrate 3--5~mm into the sample interior over timescales of 0.5--1.0~ms.

The lower panel of Figure~\ref{Fig7} displays the corresponding instantaneous velocities. The initial avalanche velocities range from 15--25~m/s, subsequently decreasing to 5--10~m/s as the avalanche progresses. The average propagation velocity across all observed events is approximately 10--20~m/s, in good agreement with the bulk NbTi data (11~m/s) reported in \cite{Ch1}. These velocities are orders of 
magnitude lower than the 14--25~km/s observed in thin-film superconductors \cite{Vestgarden, Bolz, Bolz2}, reflecting the fundamentally different propagation regime in our bulk sample.

Remarkably, when we normalize both the velocity and distance for each avalanche by their respective maximum values, all six events collapse onto a universal curve (inset of Figure~\ref{Fig7}). This normalized velocity decreases monotonically with normalized distance, dropping from $\approx$ 1.0 at the initiation point to $\approx$ 0.2--0.4 as the avalanche approaches its maximum extent. This universal behavior suggests that despite variations in individual avalanche characteristics, the underlying deceleration mechanism remains consistent across events, similar to universal scaling observed in thin-film superconductors \cite{Vestgarden, Vestgarden2}. We attribute this systematic behavior to thermally-limited propagation, as discussed in detail in Section~\ref{velocity} through analysis of characteristic timescales.

\subsection{Temperature dependence of the threshold field for the first instability}

To determine the threshold field for avalanche nucleation as a function of temperature, we performed the following procedure: A 60 mT magnetic field was applied to the sample at maximum ramp rate (for details on field establishment in the cryostat, see Section~\ref{Ramp}) at various temperatures in the range where avalanches are observable. The field establishment process was captured using a high-resolution Hamamatsu camera. By analyzing the recorded image sequences and correlating them with the time-dependent applied field, we extracted $H_{\text{th}}(T)$, the temperature-dependent threshold field at which the first avalanche appears in the camera's field of view.

\begin{figure}
\includegraphics[width=8.5cm]{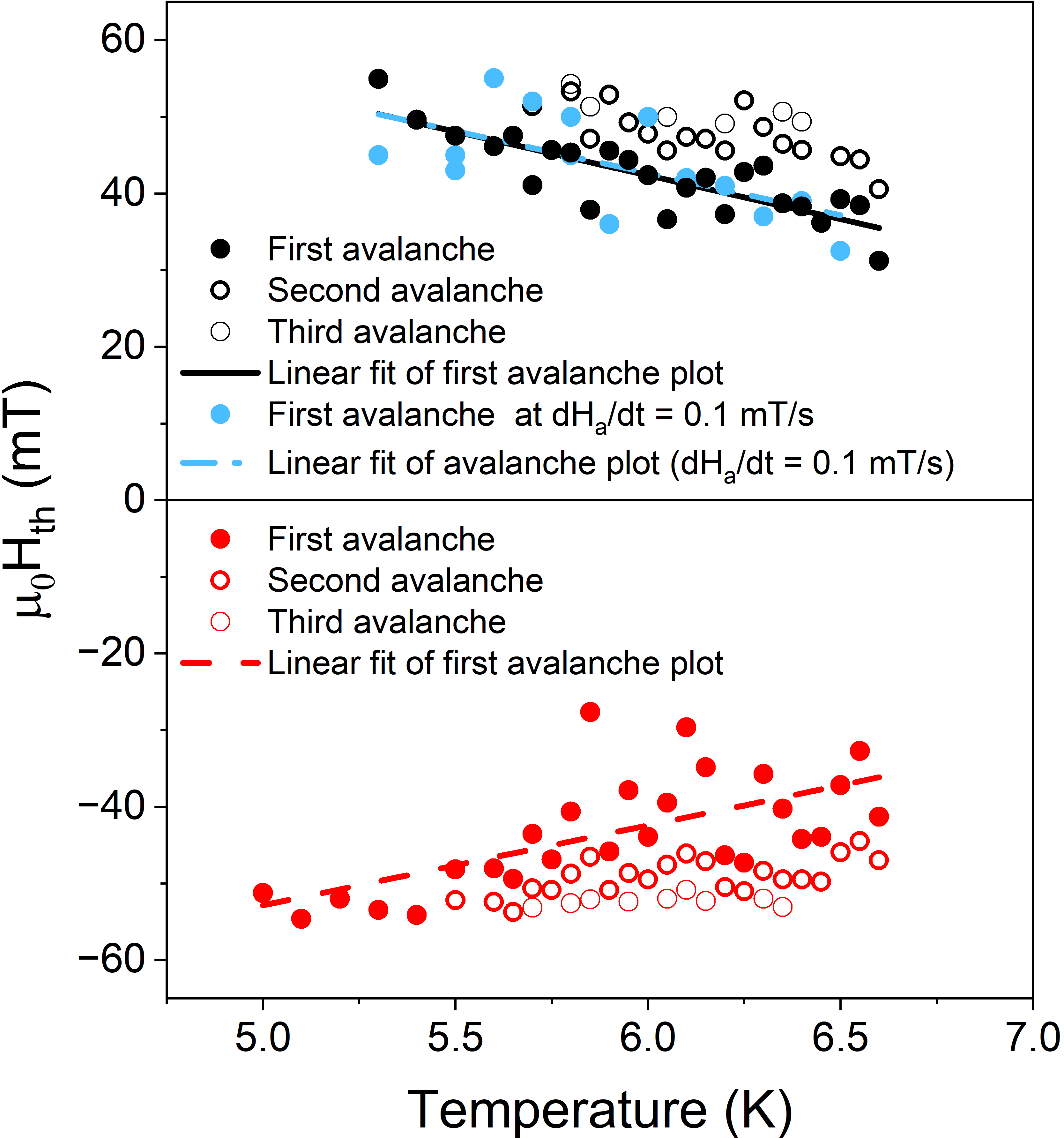}
\caption{Threshold field for avalanche triggering versus temperature. The plot displays the threshold fields at which avalanches first appear in the camera's field of view. Solid circles indicate $H_{\text{th}}$ of the first avalanche, thick and thin rings correspond to the fields at which the second and third avalanches are triggered, respectively. Black and red symbols represent the triggering fields of avalanches observed during maximum-rate field ramping in the positive and negative directions, respectively. Blue symbols correspond to the $H_{\text{th}}$ observed during slow field sweeps at $dH_{\text{a}}/dt$ = 0.1~mT/s.}
\label{Fig8}
\end{figure}

To assess the influence of field sweep rate, we compared $H_{\text{th}}(T)$ obtained under rapid field application with measurements performed using slow field ramping ($dH_{\text{a}}/dt$ = 0.1~mT/s). The measurement results are shown in Figure~\ref{Fig8}. We found that the $H_{\text{th}}(T)$ dependences measured during slow ramping and fast field application are practically identical across the entire temperature range. This independence from the sweep rate, despite a difference of several orders of magnitude in $dH_{\text{a}}/dt$, confirms that all measurements were performed under nearly adiabatic conditions, indicating that the system has sufficient time to thermalize as the external field changes. 

We observed that $H_{\text{th}}(T)$ decreases with increasing temperature, meaning that avalanches are triggered more easily at higher temperatures in the NbTi disk. Interestingly, a linear fit of $H_{\text{th}}(T)$ for the first avalanche extrapolates to zero near 9.7~K, which happens to be close to the superconducting critical temperature of the NbTi disk (9.6~K) reported in \cite{Ch5}. However, this extrapolation should be interpreted with caution, as the avalanche regime terminates above $\sim$ 6.7~K, making linear extrapolation to higher temperatures physically questionable. This behavior of $H_{\text{th}}(T)$ differs fundamentally from that observed in thin-film samples, where $H_{\text{th}}(T)$ increases monotonically with temperature and exhibits asymptotic growth approaching a threshold temperature, above which thermomagnetic avalanches are no longer observed \cite{Denisov, Vestgarden, Abaloszewa}. The reasons for this discrepancy are discussed in detail below.

\subsection{The effect of avalanche entry on the total magnetic moment of the sample}
\label{Total}

\begin{figure}
\includegraphics[width=8.5cm]{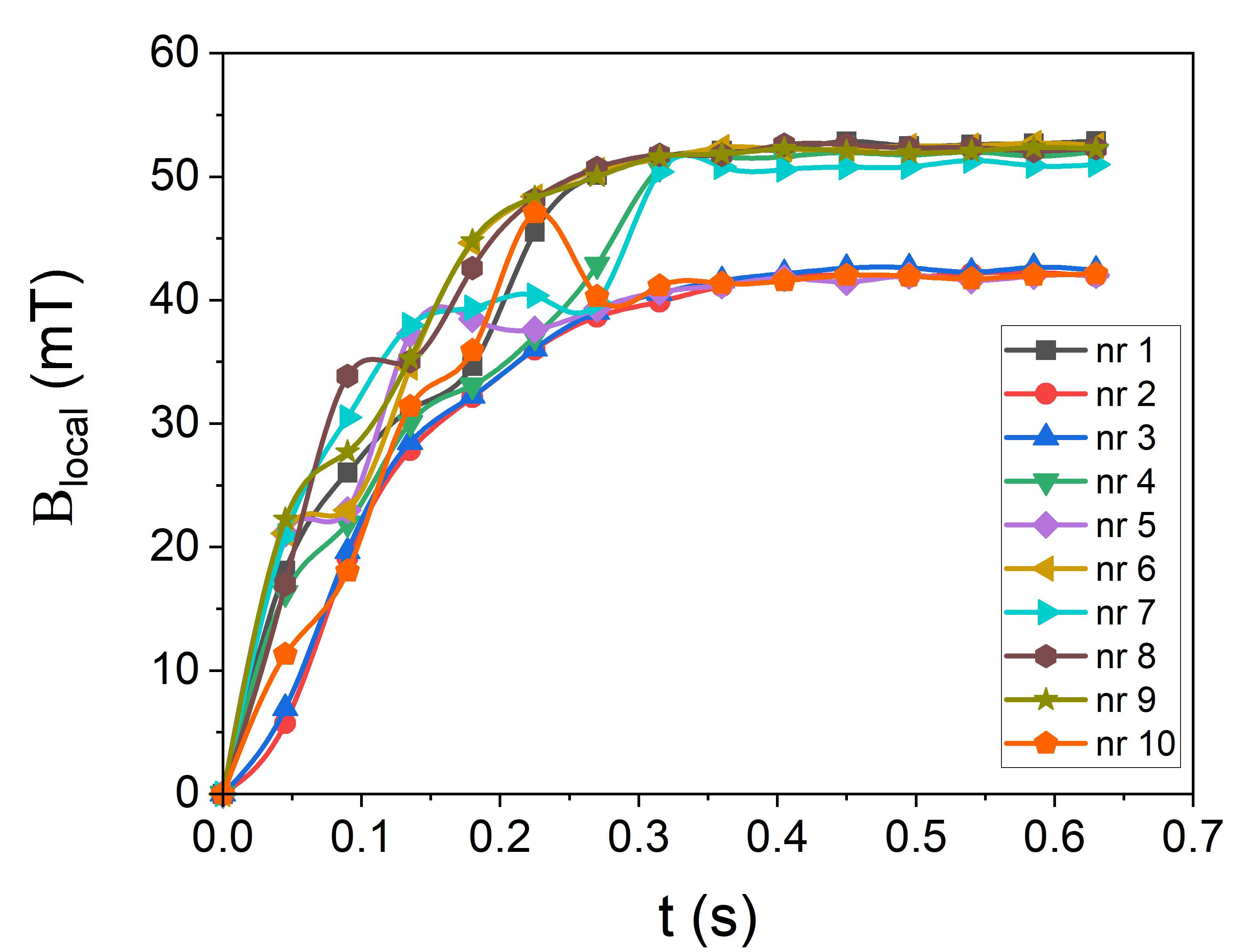}
\caption{Time evolution of the local magnetic induction measured at a point located outside the sample, at a distance of 1.5~mm from the edge. Ten measurements were performed under identical conditions (zero-field cooling followed by application of a 50~mT field at 6.5~K).}
\label{Fig9}
\end{figure}

We performed ten measurements of the local magnetic induction in the stray field near the disk, at a point 1.5~mm outside the sample edge. Each measurement followed zero-field cooling to 6.5~K and subsequent application of a 50~mT external field at the maximum rate provided by our setup. The results, summarized in Figure~\ref{Fig9}, show the temporal evolution of the induction for these ten runs. At this temperature, during such an increase in the external field, avalanches enter the sample. During each avalanche, the local magnetic induction exhibits a transient change followed by relaxation. This short-lived variation can be attributed to the redistribution of screening currents and transient overcompensation at the sample edge. As vortices enter, the demagnetizing field readjusts to restore equilibrium, briefly modifying the magnetic field outside the sample before it relaxes. The variations between measurements reflect the stochastic nature of avalanche dynamics: the timing, entry point, and propagation path all exhibit statistical variability.

The sample exhibits distinct magnetic-moment states depending on how many large avalanches have occurred. Upon field application, screening currents create the sample's magnetic moment. When an avalanche occurs, vortices previously accumulated near the edge rapidly enter the interior, reducing the vortex density outside and weakening the local magnetic field. This redistribution causes a drop in the magnetic moment, observed as a discrete jump in the magnetization signal. The jump magnitude scales with the number of vortices entering during the avalanche - larger avalanches produce larger drops. Some avalanches may occur outside the camera's field of view; however, by repositioning the sample, we observe additional events that bring substantial flux into other regions of the sample. Let us roughly estimate the avalanches area if we observe a jump between two magnetic moment "states" manifesting as a 5--10~mT change in the local magnetic flux density (Figure~\ref{Fig9}). Assuming the flux density changes by 5--10~mT in a 3~mm ring around the disk (area $\sim$113~mm$^2$), the flux change is $\Delta\Phi$ = 0.57--1.13$\times10^{-6}$~Wb, corresponding to $\sim2.8$--$5.5\times10^8$~vortices. With an average flux density of 45~mT in the avalanche region, this yields a reasonable value for the avalanche area of $\sim$13--25~mm$^2$ or about 11-22$\%$ of the disk area, which agrees well with the characteristic size of dendritic flux structures.

\section{Discussion}
\subsection{Avalanche propagation velocity}
\label{velocity}

The avalanche propagation velocity in our disk is significantly lower than in thin-film samples~\cite{Johansen_2002, Vestgarden, Aranson}. To understand the physical mechanism governing avalanche dynamics, we analyze the characteristic electromagnetic and thermal timescales in our system and compare them with experimental observations and published results for thin films.

First, we verify whether our sample can be considered electromagnetically thin in the sense used by Brandt~\cite{Brandt}. The characteristic electromagnetic penetration depth (skin depth) is:
\begin{equation}
\delta = \sqrt{\frac{2\rho_0}{\mu_0 \omega}},
\end{equation}
where $\rho_0 = 6$--$7\times10^{-7}$~$\Omega\cdot$m is the normal-state resistivity at $T \approx 10$~K~\cite{Flachbart,Hillmann}, $\mu_0$ is the permeability of free space, and $\omega \sim 1/\tau$ is the characteristic frequency. Taking a conservative lower bound of the characteristic time as $\tau \gtrsim 1~\mu$s (from magneto-optical observations, where changes in the flux-gradient pattern occur on timescales clearly exceeding $1~\mu$s) gives $\omega \lesssim 10^6$~Hz and $\delta \gtrsim 1.0$~mm. Since the disk thickness $d = 0.1$~mm $< \delta$, we have $d/\delta \lesssim 0.1 \ll 1$, confirming that our sample is indeed in the thin-disk regime for electromagnetic dynamics.

For a thin superconducting disk, the characteristic electromagnetic relaxation time is given by Brandt~\cite{Brandt}:
\begin{equation}
\tau_{\text{em}} = \frac{0.18 \mu_0 R d}{\rho_0} \approx 0.2~\mu\text{s},
\end{equation}
where $R = 6$~mm is the disk radius. The corresponding maximum electromagnetic avalanche velocity can be estimated as~\cite{Vestgarden,Vestgarden2}:
\begin{equation}
v_{\text{em}} = 0.77\frac{\rho_0}{\mu_0 d} \approx 4.3~\text{km/s}.
\end{equation}
This electromagnetic velocity is two orders of magnitude larger than our observed velocities of 15--25~m/s, indicating that electromagnetic propagation alone cannot explain the avalanche dynamics. Instead, the observed slow velocities suggest that thermal processes play the dominant role.

The thermal response of the system to heat generated by vortex motion is characterized by several material parameters. For NbTi at $T \approx 6$--$6.5$~K ($T/T_{\text{c}} \approx 0.625$), the specific heat is $C \sim 10^3$~J/(m$^3$$\cdot$K), and the thermal conductivity is $\kappa \sim 0.03$~W/(m$\cdot$K), giving a thermal diffusivity $\alpha = \kappa/C \sim 3 \times 10^{-5}$~m$^2$/s~~\cite{Flachbart, Davies, Herzog}. 
The lateral thermal diffusion time across the disk thickness is:
\begin{equation}
\tau_{\kappa} = \frac{d^2}{\alpha} \approx 330~\mu\text{s}.
\end{equation}
The heat removal time through the sample-nonadecane boundary requires estimating the effective thermal boundary conductance. The thermal conductivity of solid nonadecane (C$_{19}$H$_{40}$) at cryogenic temperatures is $\kappa_{\text{n}} \approx$0.1~W/(m$\cdot$K)~\cite{nonadecane}. With a typical layer thickness $L \sim 50$--100~$\mu$m, the effective thermal boundary conductance is:
\begin{equation}
h = \frac{\kappa_{\text{n}}}{L} \sim 10^3~\text{W/(m$^2$$\cdot$K)}.
\label{eq:h_conductance}
\end{equation}
The heat removal time through the boundary is then:
\begin{equation}
\tau_{\text{h}} = \frac{C \cdot d}{h} \sim 100~\mu\text{s}.
\label{eq:tau_h}
\end{equation}
The observed avalanche duration from magneto-optical imaging (Figure~\ref{Fig7}) is:
\begin{equation}
\tau_{\text{av}} \sim 0.5\text{--}1.0~\text{ms}.
\label{eq:tau_av}
\end{equation}
This establishes the following hierarchy of characteristic times in our system:
\begin{equation}
\tau_{\text{em}} \ll \tau_{\text{h}} < \tau_\kappa < \tau_{\text{av}}.
\label{eq:hierarchy}
\end{equation}
Numerically:
\begin{equation}
0.2~\mu\text{s} \ll 100~\mu\text{s} < 330~\mu\text{s} < 500\text{--}1000~\mu\text{s}.
\label{eq:hierarchy_numerical}
\end{equation}

This hierarchy is different from that in thin superconducting films with good thermal contact to the substrate. Vestg\aa{}rden et al.~\cite{Vestgarden,Vestgarden2} studied MgB$_2$ films on sapphire substrates and found characteristic times of: $\tau_{\text{em}} \sim 1.9$~ns, $\tau_\kappa \sim 5.1$~ns for lateral diffusion through the film thickness, and - crucially - $\tau_{\text{h}} \sim 52$~ns for heat removal to the substrate. The total avalanche duration in their films was $\tau_{\text{av}} \sim 75$--80~ns. Thus, in thin films, the hierarchy is:
\begin{equation}
\tau_{\text{em}} \sim \tau_\kappa \ll \tau_{\text{h}} \lesssim \tau_{\text{av}}.
\label{eq:hierarchy_films}
\end{equation}
This hierarchy of time scales leads to two distinct regimes of thermomagnetic flux dynamics, in agreement with the general theory of thermal and electromagnetic stability of the critical state developed by Mints and Brandt~\cite{Mints1}:

Thin films (fast regime): In films with good thermal contact, $\tau_{\text{h}} \sim 50$~ns and the avalanche duration $\tau_{\text{av}} \sim 80$~ns. The avalanche propagates at near-electromagnetic speeds ($v \sim 90$~km/s initially) driven by nonlocal electrodynamics and adiabatic heating~\cite{Vestgarden,Vestgarden2,Aranson}. The front advances faster than heat can diffuse laterally, essentially ``outrunning'' the heated region. Heat removal to the substrate eventually limits the avalanche size when heat generation is balanced by removal, but this occurs after the avalanche has penetrated deeply at high speed.

Our disk (slow regime): In our system with poor thermal contact, $\tau_{\text{h}} \sim 100$~$\mu$s and lateral diffusion is even slower ($\tau_\kappa \sim 330$~$\mu$s). The avalanche cannot ``outrun'' the heated region. Instead, propagation occurs in a state of quasi-thermal equilibrium, where the front advances only as fast as heat can diffuse laterally and be removed through the nonadecane layer. The observed deceleration from 15--25~m/s to 5--10~m/s (Figure~\ref{Fig7}) reflects progressive thermal saturation as the avalanche front advances and heat accumulates in the propagation region, reducing the driving force for further vortex motion.

An interesting phenomenon appears in some magneto-optical sequences: avalanches seem to ``emerge'' from inside the sample over several consecutive frames (e.g., frames 2--5 in Figure~\ref{Fig6}, separated by $\Delta t = 45~\mu$s each). This behavior directly reflects the different electromagnetic and thermal timescales. Since $\delta > d$, the electromagnetic field can penetrate the entire disk. However, thermal diffusion is much slower: the thermal diffusion length during one frame interval is:
\begin{equation}
\ell_{\text{frame}} = \sqrt{\alpha \cdot \Delta t} \approx 37~\mu\text{m},
\label{eq:diffusion_length}
\end{equation}
which is less than half the sample thickness. If an avalanche nucleates at the bottom surface or within the sample volume, the associated heating and reduction of critical current density will reach the visible top surface only after thermal diffusion across the thickness - a delay of several frame intervals. This ``delayed appearance'' is thus not an artifact but a direct manifestation of thermally-limited dynamics.

\subsection{Temperature dependence of the threshold field}

The temperature dependence of the threshold field for flux avalanches, $H_{\text{th}}(T)$, provides essential insight into the dominant physical mechanisms. As shown in Figure~\ref{Fig8}, $H_{\text{th}}$ decreases monotonically with increasing temperature in our NbTi disk, contrary to the increasing trend observed in thin films~\cite{Vestgarden,Denisov,Abaloszewa}. Let us link this contrasting behavior to the different physical regimes identified in our analysis of characteristic timescales (Section~\ref{velocity}).

In films with good thermal contact to the substrate, the short duration of the avalanche limits the temperature rise during its propagation. Heat generated by vortex motion is rapidly transferred to the substrate, so the instability develops in an electromagnetically dominated regime. In this case, the threshold field $H_{\text{th}}$ is determined primarily by the linear stability of the critical state against small perturbations~\cite{Mints1, Mints2, Swartz, Gurevich2}. The stability criterion reflects the competition between opposing effects: as temperature increases, the specific heat $C(T)$ and thermal conductivity $\kappa(T)$ both tend to increase, enhancing the system's ability to absorb and remove heat, while the critical current density $j_{\text{c}}(T)$ decreases, reducing magnetic stability. For systems with efficient substrate cooling, the stabilizing effects of higher $C(T)$ and stronger heat removal dominate, leading to an overall increase of $H_{\text{th}}$ with temperature~\cite{Denisov}. Above a threshold temperature $T_{\text{th}}$, avalanches are completely suppressed, and the flux penetration becomes smooth and stable.

In our NbTi disk, the situation is fundamentally different. The weak thermal coupling to the substrate and slow heat removal allow extensive heating during flux motion, so the instability becomes thermally dominated. In this regime, the threshold field is governed by the onset of thermal runaway rather than by the linear stability of the critical state. The key factor controlling this process is the available temperature interval $\Delta T = T_{\text{c}} - T_0$, which defines the maximum energy that can be absorbed before superconductivity is destroyed. As the base temperature $T_0$ increases, this interval shrinks, so that less dissipated energy is needed to drive a local region above $T_{\text{c}}$, leading to easier triggering of thermal runaway and, consequently, a lower $H_{\text{th}}$. This behavior is consistent with the thermomagnetic instability framework developed in~\cite{Mints1, Mints2}, where the onset of instability is determined by the balance between Joule heating and heat removal to the bath. Other theoretical analyses~\cite{Gurevich, Gurevich2} similarly show that weak external cooling and the temperature-dependent thermophysical parameters strongly reduce the stability margin of the critical state; consequently, insufficient heat removal and a reduced temperature interval between $T_0$ and $T_{\text{c}}$ facilitate runaway heating, setting the threshold for flux avalanches in bulk-like superconductors.

The contrasting behavior of $H_{\text{th}}(T)$ in thin films and in our disk raises the question: is it possible to determine the value of thermal boundary conductance $h_{\text{c}}$ at which the nature of the temperature dependence changes? A complementary question concerns the opposite limiting case, when heat removal is negligible.

In the adiabatic limit ($h \to 0$, no heat removal), Wipf~\cite{Wipf} derived for a semi-infinite slab:
\begin{equation}
H_{\text{th}} = \frac{1}{2}\pi^{3/2} \sqrt{\frac{C(T_{\text{c}}^2 - T^2)}{T}},
\label{eq:Wipf}
\end{equation}
where $C$ is the specific heat. For NbTi in our operating temperature range the estimate based on equation~\ref{eq:Wipf} yields $H_{\text{th}} \propto \sqrt{(T_{\text{c}}^2 - T^2)/T}$, scaling from $\propto 3.7$ at $T = 5$~K to $\propto 3.1$ at $T = 6$~K. Thus, in the purely adiabatic limit, $H_{\text{th}}$ would decrease even more steeply with temperature than observed in our system, since any finite heat removal provides some stabilization.

The sign of the temperature derivative $\text{d}H_{\text{th}}/\text{d}T$ is determined by the competition between stabilizing thermal effects (increasing $C(T)$ and $\kappa(T)$, enhanced by larger $h$) and destabilizing effects (decreasing $j_{\text{c}}(T)$ and reducing temperature margin $\Delta T = T_{\text{c}} - T$). From the general stability criterion~\cite{Mints2,Swartz}, the threshold field for a system with thermal boundary conductance $h$ can be expressed as:
\begin{equation}
H_{\text{th}}^2 \propto \frac{C(T) \kappa(T) h \cdot f[j_{\text{c}}(T)]}{g[\partial j_{\text{c}}/\partial T, \Delta T]},
\label{eq:Hth_general}
\end{equation}
where $f$ and $g$ are functions of the critical current and its temperature derivatives. While a complete analysis would require taking the temperature derivative of Equation~\ref{eq:Hth_general} and solving numerically for the condition $\text{d}H_{\text{th}}/\text{d}T = 0$, we can obtain an order-of-magnitude estimate of $h_c$ through a simpler characteristic-time analysis that captures the essential physics. The key insight is to compare two fundamental timescales: the heat transfer time across the boundary $\tau_{\text{h}}$, determined as in Equation~\ref{eq:tau_h}, and the thermal runaway time - the characteristic time for the system to heat from $T$ to $T_{\text{c}}$ under Joule heating. Following \cite{Wipf, Mints2} we find:

- power dissipation per unit volume $P = E \cdot j_{\text{c}} = \rho_{\text{ff}}\cdot j_{\text{c}}^2$,

- energy needed to heat from $T$ to $T_{\text{c}}$: $Q = C\cdot(T_{\text{c}} - T)$ per unit volume,

- thermal runaway time: $\tau_{\text{runaway}} = Q/P$.

We obtain:

\begin{equation}
\tau_{\text{runaway}} = \frac{C \cdot (T_{\text{c}} - T)}{\rho_{\text{ff}}\cdot j_{\text{c}}^2}.
\label{eq:tau_runaway}
\end{equation}

Balancing $\tau_{\text{h}}$ and $\tau_{\text{runaway}}$ and solving for $h_{\text{c}}$ we find:

\begin{equation}
h_{\text{c}} = \frac{\rho_{\text{ff}}\cdot j_{\text{c}}^2\cdot d}{T_{\text{c}} - T}.
\label{eq:h_c}
\end{equation}
Using the flux-flow resistivity $\rho_{\text{ff}} \approx \rho_{\text{n}}\cdot(B/B_{\text{c2}})$ and in the critical state near threshold $B \sim \mu_0 H_{\text{th}} \sim 0.06$~T, $B_{\text{c2}} \sim 10$ T at 6~K, $\rho_{\text{n}} \approx 4.5\cdot10^{-7} \Omega \cdot$m (lower bound of normal state resistivity \cite{Hillmann}), $\rho_{\text{ff}} \approx \rho_{\text{n}} (0.06/10) \approx 2.7\cdot10^{-9} \Omega \cdot $m, $j_{\text{c}} \approx 1.2\cdot10^{9}$~A/m$^2$, $(T_{\text{c}} - T) = 4$~K, we obtain $h_{\text{c}}\approx9.7\cdot10^{4}$ W/(m$^2\cdot$K).

Order of magnitude of $h_{\text{c}}$ is physically reasonable, lying between our experimental value ($h \sim 10^3$~W/(m$^2\cdot$K)) and typical film-substrate interfaces ($h \sim 10^5-10^6$~W/(m$^2\cdot$K) \cite{Vestgarden, Vestgarden2}).

This estimate places our system with nonadecane firmly in the thermally-limited regime where $\text{d}H_{\text{th}}/\text{d}T < 0$. Intermediate thermal coupling near $h_{\text{c}}$ could exhibit non-monotonic $H_{\text{th}}(T)$ behavior with a maximum at intermediate temperatures where the competing effects balance. Our estimate of $h_{\text{c}}$ can be tested experimentally by systematically varying the thermal boundary conductance.

Interestingly, both thin films and our disk exhibit avalanche-free behavior above a threshold temperature, but through fundamentally different mechanisms. In thin films with good thermal contact, electromagnetic stabilization above a threshold temperature $T_{\text{th}}$ suppresses avalanche nucleation while the critical current density remains sufficiently large - creating a true stability window where the system is both avalanche-free and retains strong flux pinning~\cite{Vestgarden,Denisov}. This stabilization arises from the increasing heat capacity $C(T)$ combined with fast heat removal ($\tau_{\text{h}} \sim 50$~ns), which prevents the positive thermal feedback from developing.

In our disk with weaker thermal contact, avalanche suppression above $T \sim 6.7$~K results from the breakdown of the thermomagnetic instability mechanism. The key difference lies in the operating reduced temperatures: thin films achieve stabilization at $T/T_{\text{c}} \sim 0.2$ where $j_{\text{c}}$ retains $\sim$95\% of $j_{\text{c}}(0)$~\cite{Baziljevich, Johansen_2002, Rudnev1, Rudnev2}, whereas avalanche suppression in our disk occurs at $T/T_{\text{c}} \sim 0.67$ where $j_{\text{c}}$ has fallen to $\sim$30--55\% of $j_{\text{c}}(0)$ (using $j_{\text{c}} \propto [1-(T/T_{\text{c}})^2]^n$ with $n \sim 1$--2 \cite{Kramer}). The resulting Joule heating rate ($\propto j_{\text{c}}^2$) in our disk is thus reduced by a factor of 3--10 compared to the stabilized thin film regime. Simultaneously, the increased $C(T)$ further suppresses thermal runaway. However, this avalanche suppression does not indicate enhanced stability - instead, the system enters a thermally-degraded regime with weak pinning where flux penetrates smoothly but extensively, compromising functionality.

Both the propagation mechanism (Section~\ref{velocity}) and the temperature dependence of the threshold field are determined by the same underlying factor: the thermal boundary conductance $h$ and the resulting hierarchy of thermal timescales. This unified picture highlights the critical importance of thermal management in superconducting applications and demonstrates that our NbTi disk operates in a previously uncharacterized, thermally-limited avalanche regime.

The observed magnetic moment jumps upon avalanche entry (Section~\ref{Total}) further illustrate the global consequences of local thermal instabilities, linking dendritic avalanche events to bulk-scale flux reconfiguration. At higher $T$, where $H_{\text{th}}$ decreases (Figure~\ref{Fig8}), we anticipate larger jumps due to easier triggering, weaker pinning, and amplified thermal feedback, potentially exacerbating quench risks in bulk magnets. These findings extend prior inductive studies~\cite{Ch1,Ch2,Nabialek} by linking local MOI observations to bulk magnetic response.

\section{Conclusion}

Our study provides the first systematic characterization of flux avalanche dynamics in bulk NbTi operating in a thermally-limited regime. While flux jumps in bulk superconductors have been studied since the 1960s through magneto-optical techniques~\cite{Wertheimer,Harrison}, and thin-film avalanche dynamics are well-characterized in the electromagnetically-controlled regime~\cite{Vestgarden,Denisov}, the intermediate regime of controlled poor thermal coupling has remained unexplored. The avalanches observed by us in bulk NbTi exhibit a variety of curved and dendritic morphologies and propagate at initial velocities of 15--25~m/s, systematically decelerating to 5--10~m/s as they advance into the sample - orders of magnitude slower than the 14--25~km/s observed in thin films. All avalanches exhibit universal normalized velocity-distance behavior, confirming thermally-limited propagation governed by progressive heat accumulation. Analysis of characteristic timescales shows that avalanche dynamics in bulk NbTi are thermally limited rather than purely electromagnetic in nature.

Three key advances distinguish this work from previous investigations. First, we provide the first systematic measurement of the temperature dependence of the threshold field $H_{\text{th}}(T)$ in a thermally-limited regime, demonstrating the inverse temperature dependence ($\mathrm{d}H_{\text{th}}/ \mathrm{d}T < 0$) predicted by Wipf~\cite{Wipf} for the adiabatic limit but never before confirmed through spatiotemporal imaging. While early studies explored damping effects qualitatively~\cite{Harrison}, the physical origin of different avalanche regimes remained unclear. Second, we estimate the critical thermal boundary conductance $h_{\text{c}}$ that separates thermally-limited from electromagnetically-controlled regimes, establishing thermal coupling as the fundamental control parameter. Third, we establish a quantitative hierarchy of characteristic timescales that determines the dominant physics, demonstrating that both avalanche propagation velocities and threshold field temperature dependences are manifestations of the same underlying thermal control mechanism.

The millisecond duration of thermally-limited avalanches, while problematic for device stability, enables direct real-time observation of flux dynamics without ultrafast techniques, opening new experimental opportunities for studying intermediate avalanche stages inaccessible in thin films. 

On a broader level, our results link microscopic flux redistribution to global magnetic-moment jumps and quench-like behavior, emphasizing the need to control thermal management in bulk superconductors. These findings not only deepen the understanding of thermomagnetic instabilities but also provide practical guidance for improving flux stability and quench protection in NbTi-based magnets and cryogenic current-carrying devices. By bridging the gap between thin-film and bulk superconductors, these results establish a unified view of thermomagnetic avalanche dynamics across geometries and material classes.

\section*{Acknowledgments}

This work was partially supported by the PAN-NANU 2025 Program for Long-Term Research Stays of the Polish Academy of Sciences.

\section*{Data Availability}

The data that support the findings of this study are available from the corresponding author upon reasonable request.


\begin{thebibliography}{42}%
\makeatletter
\providecommand \@ifxundefined [1]{%
 \@ifx{#1\undefined}
}%
\providecommand \@ifnum [1]{%
 \ifnum #1\expandafter \@firstoftwo
 \else \expandafter \@secondoftwo
 \fi
}%
\providecommand \@ifx [1]{%
 \ifx #1\expandafter \@firstoftwo
 \else \expandafter \@secondoftwo
 \fi
}%
\providecommand \natexlab [1]{#1}%
\providecommand \enquote  [1]{``#1''}%
\providecommand \bibnamefont  [1]{#1}%
\providecommand \bibfnamefont [1]{#1}%
\providecommand \citenamefont [1]{#1}%
\providecommand \href@noop [0]{\@secondoftwo}%
\providecommand \href [0]{\begingroup \@sanitize@url \@href}%
\providecommand \@href[1]{\@@startlink{#1}\@@href}%
\providecommand \@@href[1]{\endgroup#1\@@endlink}%
\providecommand \@sanitize@url [0]{\catcode `\\12\catcode `\$12\catcode
  `\&12\catcode `\#12\catcode `\^12\catcode `\_12\catcode `\%12\relax}%
\providecommand \@@startlink[1]{}%
\providecommand \@@endlink[0]{}%
\providecommand \url  [0]{\begingroup\@sanitize@url \@url }%
\providecommand \@url [1]{\endgroup\@href {#1}{\urlprefix }}%
\providecommand \urlprefix  [0]{URL }%
\providecommand \Eprint [0]{\href }%
\providecommand \doibase [0]{https://doi.org/}%
\providecommand \selectlanguage [0]{\@gobble}%
\providecommand \bibinfo  [0]{\@secondoftwo}%
\providecommand \bibfield  [0]{\@secondoftwo}%
\providecommand \translation [1]{[#1]}%
\providecommand \BibitemOpen [0]{}%
\providecommand \bibitemStop [0]{}%
\providecommand \bibitemNoStop [0]{.\EOS\space}%
\providecommand \EOS [0]{\spacefactor3000\relax}%
\providecommand \BibitemShut  [1]{\csname bibitem#1\endcsname}%
\let\auto@bib@innerbib\@empty
\bibitem [{\citenamefont {Baziljevich}\ \emph {et~al.}(2014)\citenamefont
  {Baziljevich}, \citenamefont {Baruch-El}, \citenamefont {Johansen},\ and\
  \citenamefont {Yeshurun}}]{Baziljevich}%
  \BibitemOpen
  \bibfield  {author} {\bibinfo {author} {\bibfnamefont {M.}~\bibnamefont
  {Baziljevich}}, \bibinfo {author} {\bibfnamefont {E.}~\bibnamefont
  {Baruch-El}}, \bibinfo {author} {\bibfnamefont {T.~H.}\ \bibnamefont
  {Johansen}},\ and\ \bibinfo {author} {\bibfnamefont {Y.}~\bibnamefont
  {Yeshurun}},\ }\bibfield  {title} {\bibinfo {title} {Dendritic instability in
  {YBa$_2$Cu$_3$O$_{7-\delta}$} films triggered by transient magnetic fields},\
  }\href {https://doi.org/10.1063/1.4887374} {\bibfield  {journal} {\bibinfo
  {journal} {Applied Physics Letters}\ }\textbf {\bibinfo {volume} {105}},\
  \bibinfo {pages} {012602} (\bibinfo {year} {2014})}\BibitemShut {NoStop}%
\bibitem [{\citenamefont {Johansen}\ \emph {et~al.}(2002)\citenamefont
  {Johansen}, \citenamefont {Baziljevich}, \citenamefont {Shantsev},
  \citenamefont {Goa}, \citenamefont {Galperin}, \citenamefont {Kang},
  \citenamefont {Kim}, \citenamefont {Choi}, \citenamefont {Kim},\ and\
  \citenamefont {Lee}}]{Johansen_2002}%
  \BibitemOpen
  \bibfield  {author} {\bibinfo {author} {\bibfnamefont {T.~H.}\ \bibnamefont
  {Johansen}}, \bibinfo {author} {\bibfnamefont {M.}~\bibnamefont
  {Baziljevich}}, \bibinfo {author} {\bibfnamefont {D.~V.}\ \bibnamefont
  {Shantsev}}, \bibinfo {author} {\bibfnamefont {P.~E.}\ \bibnamefont {Goa}},
  \bibinfo {author} {\bibfnamefont {Y.~M.}\ \bibnamefont {Galperin}}, \bibinfo
  {author} {\bibfnamefont {W.~N.}\ \bibnamefont {Kang}}, \bibinfo {author}
  {\bibfnamefont {H.~J.}\ \bibnamefont {Kim}}, \bibinfo {author} {\bibfnamefont
  {E.~M.}\ \bibnamefont {Choi}}, \bibinfo {author} {\bibfnamefont {M.-S.}\
  \bibnamefont {Kim}},\ and\ \bibinfo {author} {\bibfnamefont {S.~I.}\
  \bibnamefont {Lee}},\ }\bibfield  {title} {\bibinfo {title} {Dendritic
  magnetic instability in superconducting {MgB$_2$} films},\ }\href
  {https://doi.org/10.1209/epl/i2002-00146-1} {\bibfield  {journal} {\bibinfo
  {journal} {Europhysics Letters}\ }\textbf {\bibinfo {volume} {59}},\ \bibinfo
  {pages} {599} (\bibinfo {year} {2002})}\BibitemShut {NoStop}%
\bibitem [{\citenamefont {Awad}\ \emph {et~al.}(2011)\citenamefont {Awad},
  \citenamefont {Aliev}, \citenamefont {Ataklti}, \citenamefont {Silhanek},
  \citenamefont {Moshchalkov}, \citenamefont {Galperin},\ and\ \citenamefont
  {Vinokur}}]{Awad}%
  \BibitemOpen
  \bibfield  {author} {\bibinfo {author} {\bibfnamefont {A.~A.}\ \bibnamefont
  {Awad}}, \bibinfo {author} {\bibfnamefont {F.~G.}\ \bibnamefont {Aliev}},
  \bibinfo {author} {\bibfnamefont {G.~W.}\ \bibnamefont {Ataklti}}, \bibinfo
  {author} {\bibfnamefont {A.}~\bibnamefont {Silhanek}}, \bibinfo {author}
  {\bibfnamefont {V.~V.}\ \bibnamefont {Moshchalkov}}, \bibinfo {author}
  {\bibfnamefont {Y.~M.}\ \bibnamefont {Galperin}},\ and\ \bibinfo {author}
  {\bibfnamefont {V.}~\bibnamefont {Vinokur}},\ }\bibfield  {title} {\bibinfo
  {title} {Flux avalanches triggered by microwave depinning of magnetic
  vortices in {Pb} superconducting films},\ }\href
  {https://doi.org/10.1103/PhysRevB.84.224511} {\bibfield  {journal} {\bibinfo
  {journal} {Phys. Rev. B}\ }\textbf {\bibinfo {volume} {84}},\ \bibinfo
  {pages} {224511} (\bibinfo {year} {2011})}\BibitemShut {NoStop}%
\bibitem [{\citenamefont {Wertheimer}\ and\ \citenamefont {{le G.
  Gilchrist}}(1967)}]{Wertheimer}%
  \BibitemOpen
  \bibfield  {author} {\bibinfo {author} {\bibfnamefont {M.}~\bibnamefont
  {Wertheimer}}\ and\ \bibinfo {author} {\bibfnamefont {J.}~\bibnamefont {{le
  G. Gilchrist}}},\ }\bibfield  {title} {\bibinfo {title} {Flux jumps in type
  {II} superconductors},\ }\href
  {https://doi.org/https://doi.org/10.1016/0022-3697(67)90038-8} {\bibfield
  {journal} {\bibinfo  {journal} {Journal of Physics and Chemistry of Solids}\
  }\textbf {\bibinfo {volume} {28}},\ \bibinfo {pages} {2509} (\bibinfo {year}
  {1967})}\BibitemShut {NoStop}%
\bibitem [{\citenamefont {Rudnev}\ \emph {et~al.}(2003)\citenamefont {Rudnev},
  \citenamefont {Antonenko}, \citenamefont {Shantsev}, \citenamefont
  {Johansen},\ and\ \citenamefont {Primenko}}]{Rudnev1}%
  \BibitemOpen
  \bibfield  {author} {\bibinfo {author} {\bibfnamefont {I.}~\bibnamefont
  {Rudnev}}, \bibinfo {author} {\bibfnamefont {S.}~\bibnamefont {Antonenko}},
  \bibinfo {author} {\bibfnamefont {D.}~\bibnamefont {Shantsev}}, \bibinfo
  {author} {\bibfnamefont {T.}~\bibnamefont {Johansen}},\ and\ \bibinfo
  {author} {\bibfnamefont {A.}~\bibnamefont {Primenko}},\ }\bibfield  {title}
  {\bibinfo {title} {Dendritic flux avalanches in superconducting {Nb$_3$Sn}
  films},\ }\href
  {https://doi.org/https://doi.org/10.1016/S0011-2275(03)00157-7} {\bibfield
  {journal} {\bibinfo  {journal} {Cryogenics}\ }\textbf {\bibinfo {volume}
  {43}},\ \bibinfo {pages} {663} (\bibinfo {year} {2003})}\BibitemShut
  {NoStop}%
\bibitem [{\citenamefont {Rudnev}\ \emph {et~al.}(2005)\citenamefont {Rudnev},
  \citenamefont {Shantsev}, \citenamefont {Johansen},\ and\ \citenamefont
  {Primenko}}]{Rudnev2}%
  \BibitemOpen
  \bibfield  {author} {\bibinfo {author} {\bibfnamefont {I.~A.}\ \bibnamefont
  {Rudnev}}, \bibinfo {author} {\bibfnamefont {D.~V.}\ \bibnamefont
  {Shantsev}}, \bibinfo {author} {\bibfnamefont {T.~H.}\ \bibnamefont
  {Johansen}},\ and\ \bibinfo {author} {\bibfnamefont {A.~E.}\ \bibnamefont
  {Primenko}},\ }\bibfield  {title} {\bibinfo {title} {Avalanche-driven fractal
  flux distributions in {NbN} superconducting films},\ }\href
  {https://doi.org/10.1063/1.1992673} {\bibfield  {journal} {\bibinfo
  {journal} {Applied Physics Letters}\ }\textbf {\bibinfo {volume} {87}},\
  \bibinfo {pages} {042502} (\bibinfo {year} {2005})}\BibitemShut {NoStop}%
\bibitem [{\citenamefont {Vasiliev}\ \emph {et~al.}(2006)\citenamefont
  {Vasiliev}, \citenamefont {Nabiałek}, \citenamefont {Chabanenko},
  \citenamefont {Rusakov}, \citenamefont {Plechota},\ and\ \citenamefont
  {Szymczak}}]{Vasiliev}%
  \BibitemOpen
  \bibfield  {author} {\bibinfo {author} {\bibfnamefont {S.}~\bibnamefont
  {Vasiliev}}, \bibinfo {author} {\bibfnamefont {A.}~\bibnamefont {Nabiałek}},
  \bibinfo {author} {\bibfnamefont {V.}~\bibnamefont {Chabanenko}}, \bibinfo
  {author} {\bibfnamefont {V.}~\bibnamefont {Rusakov}}, \bibinfo {author}
  {\bibfnamefont {S.}~\bibnamefont {Plechota}},\ and\ \bibinfo {author}
  {\bibfnamefont {H.}~\bibnamefont {Szymczak}},\ }\bibfield  {title} {\bibinfo
  {title} {The structure of thermomagnetic avalanches in superconducting disc
  of {NbTi}},\ }\href {https://doi.org/10.12693/APhysPolA.109.661} {\bibfield
  {journal} {\bibinfo  {journal} {Acta Physica Polonica A}\ }\textbf {\bibinfo
  {volume} {109}},\ \bibinfo {pages} {661} (\bibinfo {year}
  {2006})}\BibitemShut {NoStop}%
\bibitem [{\citenamefont {Nulens}\ \emph {et~al.}(2023)\citenamefont {Nulens},
  \citenamefont {Lejeune}, \citenamefont {Caeyers}, \citenamefont
  {Marinković}, \citenamefont {Cools}, \citenamefont {Dausy}, \citenamefont
  {Basov}, \citenamefont {Raes}, \citenamefont {Bael}, \citenamefont {Geresdi},
  \citenamefont {Silhanek},\ and\ \citenamefont {Vondel}}]{Nulens}%
  \BibitemOpen
  \bibfield  {author} {\bibinfo {author} {\bibfnamefont {L.}~\bibnamefont
  {Nulens}}, \bibinfo {author} {\bibfnamefont {N.}~\bibnamefont {Lejeune}},
  \bibinfo {author} {\bibfnamefont {J.}~\bibnamefont {Caeyers}}, \bibinfo
  {author} {\bibfnamefont {S.}~\bibnamefont {Marinković}}, \bibinfo {author}
  {\bibfnamefont {I.}~\bibnamefont {Cools}}, \bibinfo {author} {\bibfnamefont
  {H.}~\bibnamefont {Dausy}}, \bibinfo {author} {\bibfnamefont
  {S.}~\bibnamefont {Basov}}, \bibinfo {author} {\bibfnamefont
  {B.}~\bibnamefont {Raes}}, \bibinfo {author} {\bibfnamefont {M.}~\bibnamefont
  {Bael}}, \bibinfo {author} {\bibfnamefont {A.}~\bibnamefont {Geresdi}},
  \bibinfo {author} {\bibfnamefont {A.}~\bibnamefont {Silhanek}},\ and\
  \bibinfo {author} {\bibfnamefont {J.}~\bibnamefont {Vondel}},\ }\bibfield
  {title} {\bibinfo {title} {Catastrophic magnetic flux avalanches in {NbTiN}
  superconducting resonators},\ }\href
  {https://doi.org/10.1038/s42005-023-01386-8} {\bibfield  {journal} {\bibinfo
  {journal} {Communications Physics}\ }\textbf {\bibinfo {volume} {6}}
  (\bibinfo {year} {2023})}\BibitemShut {NoStop}%
\bibitem [{\citenamefont {Motta}\ \emph {et~al.}(2013)\citenamefont {Motta},
  \citenamefont {Colauto}, \citenamefont {Ortiz}, \citenamefont {Fritzsche},
  \citenamefont {Cuppens}, \citenamefont {Gillijns}, \citenamefont
  {Moshchalkov}, \citenamefont {Johansen}, \citenamefont {Sanchez},\ and\
  \citenamefont {Silhanek}}]{Motta}%
  \BibitemOpen
  \bibfield  {author} {\bibinfo {author} {\bibfnamefont {M.}~\bibnamefont
  {Motta}}, \bibinfo {author} {\bibfnamefont {F.}~\bibnamefont {Colauto}},
  \bibinfo {author} {\bibfnamefont {W.}~\bibnamefont {Ortiz}}, \bibinfo
  {author} {\bibfnamefont {J.}~\bibnamefont {Fritzsche}}, \bibinfo {author}
  {\bibfnamefont {J.}~\bibnamefont {Cuppens}}, \bibinfo {author} {\bibfnamefont
  {W.}~\bibnamefont {Gillijns}}, \bibinfo {author} {\bibfnamefont
  {V.}~\bibnamefont {Moshchalkov}}, \bibinfo {author} {\bibfnamefont
  {T.}~\bibnamefont {Johansen}}, \bibinfo {author} {\bibfnamefont
  {A.}~\bibnamefont {Sanchez}},\ and\ \bibinfo {author} {\bibfnamefont
  {A.}~\bibnamefont {Silhanek}},\ }\bibfield  {title} {\bibinfo {title}
  {Enhanced pinning in superconducting thin films with graded pinning
  landscapes},\ }\href {https://doi.org/10.1063/1.4807848} {\bibfield
  {journal} {\bibinfo  {journal} {Applied Physics Letters}\ }\textbf {\bibinfo
  {volume} {102}},\ \bibinfo {pages} {212601} (\bibinfo {year}
  {2013})}\BibitemShut {NoStop}%
\bibitem [{\citenamefont {Wimbush}\ \emph {et~al.}(2004)\citenamefont
  {Wimbush}, \citenamefont {Holzapfel},\ and\ \citenamefont {Jooss}}]{Wimbush}%
  \BibitemOpen
  \bibfield  {author} {\bibinfo {author} {\bibfnamefont {S.~C.}\ \bibnamefont
  {Wimbush}}, \bibinfo {author} {\bibfnamefont {B.}~\bibnamefont {Holzapfel}},\
  and\ \bibinfo {author} {\bibfnamefont {C.}~\bibnamefont {Jooss}},\ }\bibfield
   {title} {\bibinfo {title} {Observation of dendritic flux instabilities in
  {YNi$_2$B$_2$C} thin films},\ }\href {https://doi.org/10.1063/1.1778816}
  {\bibfield  {journal} {\bibinfo  {journal} {Journal of Applied Physics}\
  }\textbf {\bibinfo {volume} {96}},\ \bibinfo {pages} {3589} (\bibinfo {year}
  {2004})}\BibitemShut {NoStop}%
\bibitem [{\citenamefont {Behnia}\ \emph {et~al.}(2000)\citenamefont {Behnia},
  \citenamefont {Capan}, \citenamefont {Mailly},\ and\ \citenamefont
  {Etienne}}]{Behnia}%
  \BibitemOpen
  \bibfield  {author} {\bibinfo {author} {\bibfnamefont {K.}~\bibnamefont
  {Behnia}}, \bibinfo {author} {\bibfnamefont {C.}~\bibnamefont {Capan}},
  \bibinfo {author} {\bibfnamefont {D.}~\bibnamefont {Mailly}},\ and\ \bibinfo
  {author} {\bibfnamefont {B.}~\bibnamefont {Etienne}},\ }\bibfield  {title}
  {\bibinfo {title} {Internal avalanches in a pile of superconducting
  vortices},\ }\href {https://doi.org/10.1103/PhysRevB.61.R3815} {\bibfield
  {journal} {\bibinfo  {journal} {Phys. Rev. B}\ }\textbf {\bibinfo {volume}
  {61}},\ \bibinfo {pages} {R3815} (\bibinfo {year} {2000})}\BibitemShut
  {NoStop}%
\bibitem [{\citenamefont {Harrison}\ \emph {et~al.}(1973)\citenamefont
  {Harrison}, \citenamefont {Wright},\ and\ \citenamefont
  {Wertheimer}}]{Harrison}%
  \BibitemOpen
  \bibfield  {author} {\bibinfo {author} {\bibfnamefont {R.~B.}\ \bibnamefont
  {Harrison}}, \bibinfo {author} {\bibfnamefont {L.~S.}\ \bibnamefont
  {Wright}},\ and\ \bibinfo {author} {\bibfnamefont {M.~R.}\ \bibnamefont
  {Wertheimer}},\ }\bibfield  {title} {\bibinfo {title} {Kinetics of flux jumps
  in {type-II} superconductors},\ }\href
  {https://doi.org/10.1103/PhysRevB.7.1864} {\bibfield  {journal} {\bibinfo
  {journal} {Phys. Rev. B}\ }\textbf {\bibinfo {volume} {7}},\ \bibinfo {pages}
  {1864} (\bibinfo {year} {1973})}\BibitemShut {NoStop}%
\bibitem [{\citenamefont {Chabanenko}\ \emph {et~al.}(2022)\citenamefont
  {Chabanenko}, \citenamefont {Nabiałek},\ and\ \citenamefont
  {Puźniak}}]{Ch1}%
  \BibitemOpen
  \bibfield  {author} {\bibinfo {author} {\bibfnamefont {V.}~\bibnamefont
  {Chabanenko}}, \bibinfo {author} {\bibfnamefont {A.}~\bibnamefont
  {Nabiałek}},\ and\ \bibinfo {author} {\bibfnamefont {R.}~\bibnamefont
  {Puźniak}},\ }\bibfield  {title} {\bibinfo {title} {Multi-steps magnetic
  flux entrance/exit at thermomagnetic avalanches in the plates of hard
  superconductors},\ }\href {https://doi.org/10.3390/ma15062037} {\bibfield
  {journal} {\bibinfo  {journal} {Materials}\ }\textbf {\bibinfo {volume}
  {15}},\ \bibinfo {pages} {2037} (\bibinfo {year} {2022})}\BibitemShut
  {NoStop}%
\bibitem [{\citenamefont {Chabanenko}\ \emph {et~al.}(2003)\citenamefont
  {Chabanenko}, \citenamefont {Puźniak}, \citenamefont {Nabiałek},
  \citenamefont {Vasiliev}, \citenamefont {Rusakov}, \citenamefont {Huanqian},
  \citenamefont {Szymczak}, \citenamefont {Szymczak}, \citenamefont {Jun},
  \citenamefont {Janusz},\ and\ \citenamefont {Finkel}}]{Ch2}%
  \BibitemOpen
  \bibfield  {author} {\bibinfo {author} {\bibfnamefont {V.}~\bibnamefont
  {Chabanenko}}, \bibinfo {author} {\bibfnamefont {R.}~\bibnamefont
  {Puźniak}}, \bibinfo {author} {\bibfnamefont {A.}~\bibnamefont {Nabiałek}},
  \bibinfo {author} {\bibfnamefont {S.}~\bibnamefont {Vasiliev}}, \bibinfo
  {author} {\bibfnamefont {V.}~\bibnamefont {Rusakov}}, \bibinfo {author}
  {\bibfnamefont {L.}~\bibnamefont {Huanqian}}, \bibinfo {author}
  {\bibfnamefont {R.}~\bibnamefont {Szymczak}}, \bibinfo {author}
  {\bibfnamefont {H.}~\bibnamefont {Szymczak}}, \bibinfo {author}
  {\bibfnamefont {J.}~\bibnamefont {Jun}}, \bibinfo {author} {\bibfnamefont
  {K.}~\bibnamefont {Janusz}},\ and\ \bibinfo {author} {\bibfnamefont
  {V.}~\bibnamefont {Finkel}},\ }\bibfield  {title} {\bibinfo {title} {Flux
  jumps and {H-T} diagram of instability for {MgB$_2$}},\ }\href
  {https://doi.org/110.1023/A:1022236117354} {\bibfield  {journal} {\bibinfo
  {journal} {Journal of Low Temperature Physics}\ }\textbf {\bibinfo {volume}
  {130}},\ \bibinfo {pages} {175–191} (\bibinfo {year} {2003})}\BibitemShut
  {NoStop}%
\bibitem [{\citenamefont {Nabiałek}\ \emph {et~al.}(2003)\citenamefont
  {Nabiałek}, \citenamefont {Chabanenko}, \citenamefont {Rusakov},
  \citenamefont {Vasiliev}, \citenamefont {Szymczak}, \citenamefont {Piechota},
  \citenamefont {Dabkowska}, \citenamefont {Dabkowski}, \citenamefont {Gaulin},
  \citenamefont {Niewczas},\ and\ \citenamefont {Mironov}}]{Ch3}%
  \BibitemOpen
  \bibfield  {author} {\bibinfo {author} {\bibfnamefont {A.}~\bibnamefont
  {Nabiałek}}, \bibinfo {author} {\bibfnamefont {V.}~\bibnamefont
  {Chabanenko}}, \bibinfo {author} {\bibfnamefont {V.}~\bibnamefont {Rusakov}},
  \bibinfo {author} {\bibfnamefont {S.}~\bibnamefont {Vasiliev}}, \bibinfo
  {author} {\bibfnamefont {H.}~\bibnamefont {Szymczak}}, \bibinfo {author}
  {\bibfnamefont {S.}~\bibnamefont {Piechota}}, \bibinfo {author}
  {\bibfnamefont {H.}~\bibnamefont {Dabkowska}}, \bibinfo {author}
  {\bibfnamefont {A.}~\bibnamefont {Dabkowski}}, \bibinfo {author}
  {\bibfnamefont {B.~D.}\ \bibnamefont {Gaulin}}, \bibinfo {author}
  {\bibfnamefont {M.}~\bibnamefont {Niewczas}},\ and\ \bibinfo {author}
  {\bibfnamefont {O.}~\bibnamefont {Mironov}},\ }\bibfield  {title} {\bibinfo
  {title} {The peculiarities of magnetic flux dynamics at magnetothermal
  instability in textured {Bi$_2$Sr$_2$CaCu$_2$O$_{8+\delta}$}},\ }\href
  {https://doi.org/10.1023/A:1022212906876} {\bibfield  {journal} {\bibinfo
  {journal} {Journal of Low Temperature Physics}\ }\textbf {\bibinfo {volume}
  {130}},\ \bibinfo {pages} {425–433} (\bibinfo {year} {2003})}\BibitemShut
  {NoStop}%
\bibitem [{\citenamefont {Vestg{\aa}rden}\ \emph {et~al.}(2011)\citenamefont
  {Vestg{\aa}rden}, \citenamefont {Shantsev}, \citenamefont {Galperin},\ and\
  \citenamefont {Johansen}}]{Vestgarden}%
  \BibitemOpen
  \bibfield  {author} {\bibinfo {author} {\bibfnamefont {J.~I.}\ \bibnamefont
  {Vestg{\aa}rden}}, \bibinfo {author} {\bibfnamefont {D.~V.}\ \bibnamefont
  {Shantsev}}, \bibinfo {author} {\bibfnamefont {Y.~M.}\ \bibnamefont
  {Galperin}},\ and\ \bibinfo {author} {\bibfnamefont {T.~H.}\ \bibnamefont
  {Johansen}},\ }\bibfield  {title} {\bibinfo {title} {Dynamics and morphology
  of dendritic flux avalanches in superconducting films},\ }\href
  {https://doi.org/10.1103/PhysRevB.84.054537} {\bibfield  {journal} {\bibinfo
  {journal} {Phys. Rev. B}\ }\textbf {\bibinfo {volume} {84}},\ \bibinfo
  {pages} {054537} (\bibinfo {year} {2011})}\BibitemShut {NoStop}%
\bibitem [{\citenamefont {Bolz}(2002)}]{Bolz}%
  \BibitemOpen
  \bibfield  {author} {\bibinfo {author} {\bibfnamefont {U.}~\bibnamefont
  {Bolz}},\ }\bibfield  {title} {\bibinfo {title} {{Magnetooptische
  Untersuchungen der Flussdynamik in YBaCuO-Filmen auf ultra-kurzen
  Zeitskalen}},\ }\href@noop {} {\bibfield  {journal} {\bibinfo  {journal}
  {Ph.D. thesis (University of Konstanz, Germany)}\ } (\bibinfo {year}
  {2002})}\BibitemShut {NoStop}%
\bibitem [{\citenamefont {Bolz}\ \emph {et~al.}(2003)\citenamefont {Bolz},
  \citenamefont {Biehler}, \citenamefont {Schmidt}, \citenamefont {Runge},\
  and\ \citenamefont {Leiderer}}]{Bolz2}%
  \BibitemOpen
  \bibfield  {author} {\bibinfo {author} {\bibfnamefont {U.}~\bibnamefont
  {Bolz}}, \bibinfo {author} {\bibfnamefont {B.}~\bibnamefont {Biehler}},
  \bibinfo {author} {\bibfnamefont {D.}~\bibnamefont {Schmidt}}, \bibinfo
  {author} {\bibfnamefont {B.-U.}\ \bibnamefont {Runge}},\ and\ \bibinfo
  {author} {\bibfnamefont {P.}~\bibnamefont {Leiderer}},\ }\bibfield  {title}
  {\bibinfo {title} {Dynamics of the dendritic flux instability in
  {YBa$_2$Cu$_3$O$_{7-\delta}$} films},\ }\href
  {https://doi.org/10.1209/epl/i2003-00261-y} {\bibfield  {journal} {\bibinfo
  {journal} {Europhysics Letters}\ }\textbf {\bibinfo {volume} {64}},\ \bibinfo
  {pages} {517} (\bibinfo {year} {2003})}\BibitemShut {NoStop}%
\bibitem [{\citenamefont {Chabanenko}\ \emph {et~al.}(2024)\citenamefont
  {Chabanenko}, \citenamefont {Abaloszewa}, \citenamefont {Rusakov},
  \citenamefont {Kuchuk}, \citenamefont {Chumak}, \citenamefont {Nabiałek},
  \citenamefont {Abaloszew}, \citenamefont {Filippov},\ and\ \citenamefont
  {Puźniak}}]{Ch4}%
  \BibitemOpen
  \bibfield  {author} {\bibinfo {author} {\bibfnamefont {V.~V.}\ \bibnamefont
  {Chabanenko}}, \bibinfo {author} {\bibfnamefont {I.}~\bibnamefont
  {Abaloszewa}}, \bibinfo {author} {\bibfnamefont {V.~F.}\ \bibnamefont
  {Rusakov}}, \bibinfo {author} {\bibfnamefont {O.~I.}\ \bibnamefont {Kuchuk}},
  \bibinfo {author} {\bibfnamefont {O.~M.}\ \bibnamefont {Chumak}}, \bibinfo
  {author} {\bibfnamefont {A.}~\bibnamefont {Nabiałek}}, \bibinfo {author}
  {\bibfnamefont {A.}~\bibnamefont {Abaloszew}}, \bibinfo {author}
  {\bibfnamefont {A.}~\bibnamefont {Filippov}},\ and\ \bibinfo {author}
  {\bibfnamefont {R.}~\bibnamefont {Puźniak}},\ }\href
  {https://arxiv.org/abs/2412.19298} {\bibinfo {title} {Transformation of the
  trapped flux in a {SC} disc under electromagnetic exposure}} (\bibinfo {year}
  {2024}),\ \Eprint {https://arxiv.org/abs/2412.19298} {arXiv:2412.19298
  [cond-mat.supr-con]} \BibitemShut {NoStop}%
\bibitem [{\citenamefont {Rastgarkafshgarkolaei}\ \emph
  {et~al.}(2016)\citenamefont {Rastgarkafshgarkolaei}, \citenamefont {Zeng},\
  and\ \citenamefont {Khodadadi}}]{nonadecane}%
  \BibitemOpen
  \bibfield  {author} {\bibinfo {author} {\bibfnamefont {R.}~\bibnamefont
  {Rastgarkafshgarkolaei}}, \bibinfo {author} {\bibfnamefont {Y.}~\bibnamefont
  {Zeng}},\ and\ \bibinfo {author} {\bibfnamefont {J.~M.}\ \bibnamefont
  {Khodadadi}},\ }\bibfield  {title} {\bibinfo {title} {A molecular dynamics
  study of the effect of thermal boundary conductance on thermal transport of
  ideal crystal of n-alkanes with different number of carbon atoms},\ }\href
  {https://doi.org/10.1063/1.4952411} {\bibfield  {journal} {\bibinfo
  {journal} {Journal of Applied Physics}\ }\textbf {\bibinfo {volume} {119}},\
  \bibinfo {pages} {205107} (\bibinfo {year} {2016})}\BibitemShut {NoStop}%
\bibitem [{\citenamefont {{M{\&}I Materials Ltd.}}(2018)}]{ApiezonN}%
  \BibitemOpen
  \bibfield  {author} {\bibinfo {author} {\bibnamefont {{M{\&}I Materials
  Ltd.}}},\ }\href@noop {} {\bibinfo {title} {Apiezon {N} technical data
  sheet}},\ \bibinfo {howpublished}
  {\url{https://www.2spi.com/catalog/documents/ApiezonN (TDS).pdf}} (\bibinfo
  {year} {2018})\BibitemShut {NoStop}%
\bibitem [{\citenamefont {Calatroni}(2020)}]{Calatroni}%
  \BibitemOpen
  \bibfield  {author} {\bibinfo {author} {\bibfnamefont {S.}~\bibnamefont
  {Calatroni}},\ }\href {https://doi.org/10.48550/arXiv.2006.02842} {\bibinfo
  {title} {Materials $\&$ properties: Thermal $\&$ electrical characteristics}}
  (\bibinfo {year} {2020}),\ \Eprint {https://arxiv.org/abs/2006.02842}
  {arXiv:2006.02842 [cond-mat.supr-con]} \BibitemShut {NoStop}%
\bibitem [{\citenamefont {Sawada}\ \emph {et~al.}(1994)\citenamefont {Sawada},
  \citenamefont {Sakatsume}, \citenamefont {Goto}, \citenamefont {Nakamura},
  \citenamefont {Matsui}, \citenamefont {Settai}, \citenamefont {Ohtani},
  \citenamefont {Watanabe},\ and\ \citenamefont {Hoshi}}]{Sawada}%
  \BibitemOpen
  \bibfield  {author} {\bibinfo {author} {\bibfnamefont {A.}~\bibnamefont
  {Sawada}}, \bibinfo {author} {\bibfnamefont {S.}~\bibnamefont {Sakatsume}},
  \bibinfo {author} {\bibfnamefont {T.}~\bibnamefont {Goto}}, \bibinfo {author}
  {\bibfnamefont {S.}~\bibnamefont {Nakamura}}, \bibinfo {author}
  {\bibfnamefont {H.}~\bibnamefont {Matsui}}, \bibinfo {author} {\bibfnamefont
  {R.}~\bibnamefont {Settai}}, \bibinfo {author} {\bibfnamefont
  {Y.}~\bibnamefont {Ohtani}}, \bibinfo {author} {\bibfnamefont
  {K.}~\bibnamefont {Watanabe}},\ and\ \bibinfo {author} {\bibfnamefont
  {A.}~\bibnamefont {Hoshi}},\ }\bibfield  {title} {\bibinfo {title} {Hall
  sensor applicable to cryogenic temperatures for magnetic fields up to 25
  {T}},\ }\href {https://doi.org/https://doi.org/10.1016/0011-2275(94)90082-5}
  {\bibfield  {journal} {\bibinfo  {journal} {Cryogenics}\ }\textbf {\bibinfo
  {volume} {34}},\ \bibinfo {pages} {953} (\bibinfo {year} {1994})}\BibitemShut
  {NoStop}%
\bibitem [{\citenamefont {Baziljevich}\ \emph {et~al.}(2012)\citenamefont
  {Baziljevich}, \citenamefont {Barness}, \citenamefont {Sinvani},
  \citenamefont {Perel}, \citenamefont {Shaulov},\ and\ \citenamefont
  {Yeshurun}}]{Baziljevich2}%
  \BibitemOpen
  \bibfield  {author} {\bibinfo {author} {\bibfnamefont {M.}~\bibnamefont
  {Baziljevich}}, \bibinfo {author} {\bibfnamefont {D.}~\bibnamefont
  {Barness}}, \bibinfo {author} {\bibfnamefont {M.}~\bibnamefont {Sinvani}},
  \bibinfo {author} {\bibfnamefont {E.}~\bibnamefont {Perel}}, \bibinfo
  {author} {\bibfnamefont {A.}~\bibnamefont {Shaulov}},\ and\ \bibinfo {author}
  {\bibfnamefont {Y.}~\bibnamefont {Yeshurun}},\ }\bibfield  {title} {\bibinfo
  {title} {Magneto-optical system for high speed real time imaging},\ }\href
  {https://doi.org/10.1063/1.4746255} {\bibfield  {journal} {\bibinfo
  {journal} {Review of Scientific Instruments}\ }\textbf {\bibinfo {volume}
  {83}},\ \bibinfo {pages} {083707} (\bibinfo {year} {2012})}\BibitemShut
  {NoStop}%
\bibitem [{\citenamefont {Vestg{\aa}rden}\ \emph {et~al.}(2012)\citenamefont
  {Vestg{\aa}rden}, \citenamefont {Shantsev}, \citenamefont {Galperin},\ and\
  \citenamefont {Johansen}}]{Vestgarden2}%
  \BibitemOpen
  \bibfield  {author} {\bibinfo {author} {\bibfnamefont {J.~I.}\ \bibnamefont
  {Vestg{\aa}rden}}, \bibinfo {author} {\bibfnamefont {D.~V.}\ \bibnamefont
  {Shantsev}}, \bibinfo {author} {\bibfnamefont {Y.~M.}\ \bibnamefont
  {Galperin}},\ and\ \bibinfo {author} {\bibfnamefont {T.~H.}\ \bibnamefont
  {Johansen}},\ }\bibfield  {title} {\bibinfo {title} {Lightning in
  superconductors},\ }\href {https://doi.org/10.1038/srep00886} {\bibfield
  {journal} {\bibinfo  {journal} {Scientific Reports}\ }\textbf {\bibinfo
  {volume} {2}},\ \bibinfo {pages} {886} (\bibinfo {year} {2012})}\BibitemShut
  {NoStop}%
\bibitem [{\citenamefont {Chabanenko}\ \emph {et~al.}(2023)\citenamefont
  {Chabanenko}, \citenamefont {Nabiałek}, \citenamefont {Puźniak},\ and\
  \citenamefont {Rusakov}}]{Ch5}%
  \BibitemOpen
  \bibfield  {author} {\bibinfo {author} {\bibfnamefont {V.~V.}\ \bibnamefont
  {Chabanenko}}, \bibinfo {author} {\bibfnamefont {A.}~\bibnamefont
  {Nabiałek}}, \bibinfo {author} {\bibfnamefont {R.}~\bibnamefont
  {Puźniak}},\ and\ \bibinfo {author} {\bibfnamefont {V.~F.}\ \bibnamefont
  {Rusakov}},\ }\bibfield  {title} {\bibinfo {title} {Avalanche dynamics of
  magnetic flux in the {Nb-Ti} superconducting ring},\ }\href
  {https://doi.org/10.1088/1361-6668/acb10f} {\bibfield  {journal} {\bibinfo
  {journal} {Superconductor Science and Technology}\ }\textbf {\bibinfo
  {volume} {36}},\ \bibinfo {pages} {035010} (\bibinfo {year}
  {2023})}\BibitemShut {NoStop}%
\bibitem [{\citenamefont {Denisov}\ \emph {et~al.}(2006)\citenamefont
  {Denisov}, \citenamefont {Shantsev}, \citenamefont {Galperin}, \citenamefont
  {Choi}, \citenamefont {Lee}, \citenamefont {Lee}, \citenamefont {Bobyl},
  \citenamefont {Goa}, \citenamefont {Olsen},\ and\ \citenamefont
  {Johansen}}]{Denisov}%
  \BibitemOpen
  \bibfield  {author} {\bibinfo {author} {\bibfnamefont {D.~V.}\ \bibnamefont
  {Denisov}}, \bibinfo {author} {\bibfnamefont {D.~V.}\ \bibnamefont
  {Shantsev}}, \bibinfo {author} {\bibfnamefont {Y.~M.}\ \bibnamefont
  {Galperin}}, \bibinfo {author} {\bibfnamefont {E.-M.}\ \bibnamefont {Choi}},
  \bibinfo {author} {\bibfnamefont {H.-S.}\ \bibnamefont {Lee}}, \bibinfo
  {author} {\bibfnamefont {S.-I.}\ \bibnamefont {Lee}}, \bibinfo {author}
  {\bibfnamefont {A.~V.}\ \bibnamefont {Bobyl}}, \bibinfo {author}
  {\bibfnamefont {P.~E.}\ \bibnamefont {Goa}}, \bibinfo {author} {\bibfnamefont
  {A.~A.~F.}\ \bibnamefont {Olsen}},\ and\ \bibinfo {author} {\bibfnamefont
  {T.~H.}\ \bibnamefont {Johansen}},\ }\bibfield  {title} {\bibinfo {title}
  {Onset of dendritic flux avalanches in superconducting films},\ }\href
  {https://doi.org/10.1103/PhysRevLett.97.077002} {\bibfield  {journal}
  {\bibinfo  {journal} {Phys. Rev. Lett.}\ }\textbf {\bibinfo {volume} {97}},\
  \bibinfo {pages} {077002} (\bibinfo {year} {2006})}\BibitemShut {NoStop}%
\bibitem [{\citenamefont {Abaloszewa}\ \emph {et~al.}(2023)\citenamefont
  {Abaloszewa}, \citenamefont {Cieplak},\ and\ \citenamefont
  {Abaloszew}}]{Abaloszewa}%
  \BibitemOpen
  \bibfield  {author} {\bibinfo {author} {\bibfnamefont {I.}~\bibnamefont
  {Abaloszewa}}, \bibinfo {author} {\bibfnamefont {M.}~\bibnamefont
  {Cieplak}},\ and\ \bibinfo {author} {\bibfnamefont {A.}~\bibnamefont
  {Abaloszew}},\ }\bibfield  {title} {\bibinfo {title} {Thermomagnetic
  instabilities in {Nb} films deposited on glass substrates},\ }\href
  {https://doi.org/10.12693/APhysPolA.143.123} {\bibfield  {journal} {\bibinfo
  {journal} {Acta Physica Polonica A}\ }\textbf {\bibinfo {volume} {143}},\
  \bibinfo {pages} {123} (\bibinfo {year} {2023})}\BibitemShut {NoStop}%
\bibitem [{\citenamefont {Aranson}\ \emph {et~al.}(2001)\citenamefont
  {Aranson}, \citenamefont {Gurevich}, \citenamefont {Welling}, \citenamefont
  {Wijngaarden}, \citenamefont {Vlasko-Vlasov}, \citenamefont {Vinokur},\ and\
  \citenamefont {Welp}}]{Aranson}%
  \BibitemOpen
  \bibfield  {author} {\bibinfo {author} {\bibfnamefont {I.~S.}\ \bibnamefont
  {Aranson}}, \bibinfo {author} {\bibfnamefont {A.}~\bibnamefont {Gurevich}},
  \bibinfo {author} {\bibfnamefont {M.~S.}\ \bibnamefont {Welling}}, \bibinfo
  {author} {\bibfnamefont {R.~J.}\ \bibnamefont {Wijngaarden}}, \bibinfo
  {author} {\bibfnamefont {V.~K.}\ \bibnamefont {Vlasko-Vlasov}}, \bibinfo
  {author} {\bibfnamefont {V.~M.}\ \bibnamefont {Vinokur}},\ and\ \bibinfo
  {author} {\bibfnamefont {U.}~\bibnamefont {Welp}},\ }\bibfield  {title}
  {\bibinfo {title} {Dendritic flux avalanches and nonlocal electrodynamics in
  thin superconducting films},\ }\href
  {https://doi.org/10.1103/PhysRevLett.87.067003} {\bibfield  {journal}
  {\bibinfo  {journal} {Phys. Rev. Lett.}\ }\textbf {\bibinfo {volume} {87}},\
  \bibinfo {pages} {067003} (\bibinfo {year} {2001})}\BibitemShut {NoStop}%
\bibitem [{\citenamefont {Brandt}(1994)}]{Brandt}%
  \BibitemOpen
  \bibfield  {author} {\bibinfo {author} {\bibfnamefont {E.~H.}\ \bibnamefont
  {Brandt}},\ }\bibfield  {title} {\bibinfo {title} {Thin superconductors in a
  perpendicular magnetic ac field. {II}. circular disk},\ }\href
  {https://doi.org/10.1103/PhysRevB.50.4034} {\bibfield  {journal} {\bibinfo
  {journal} {Phys. Rev. B}\ }\textbf {\bibinfo {volume} {50}},\ \bibinfo
  {pages} {4034} (\bibinfo {year} {1994})}\BibitemShut {NoStop}%
\bibitem [{\citenamefont {Flachbart}\ \emph {et~al.}(1978)\citenamefont
  {Flachbart}, \citenamefont {Feher}, \citenamefont {J{\'{a}}no{\v{s}}},
  \citenamefont {M{\'{a}}lek},\ and\ \citenamefont {Ryska}}]{Flachbart}%
  \BibitemOpen
  \bibfield  {author} {\bibinfo {author} {\bibfnamefont {K.}~\bibnamefont
  {Flachbart}}, \bibinfo {author} {\bibfnamefont {A.}~\bibnamefont {Feher}},
  \bibinfo {author} {\bibfnamefont {{\v{S}}.}~\bibnamefont
  {J{\'{a}}no{\v{s}}}}, \bibinfo {author} {\bibfnamefont {Z.}~\bibnamefont
  {M{\'{a}}lek}},\ and\ \bibinfo {author} {\bibfnamefont {A.}~\bibnamefont
  {Ryska}},\ }\bibfield  {title} {\bibinfo {title} {Thermal conductivity of
  {NbTi} alloy in the low-temperature range},\ }\href
  {https://doi.org/https://doi.org/10.1002/pssb.2220850217} {\bibfield
  {journal} {\bibinfo  {journal} {Physica Status Solidi (b)}\ }\textbf
  {\bibinfo {volume} {85}},\ \bibinfo {pages} {545} (\bibinfo {year}
  {1978})}\BibitemShut {NoStop}%
\bibitem [{\citenamefont {Hillmann}(1981)}]{Hillmann}%
  \BibitemOpen
  \bibfield  {author} {\bibinfo {author} {\bibfnamefont {H.}~\bibnamefont
  {Hillmann}},\ }\bibinfo {title} {Fabrication technology of superconducting
  material},\ in\ \href {https://doi.org/10.1007/978-1-4757-0037-4_5} {\emph
  {\bibinfo {booktitle} {Superconductor Materials Science: Metallurgy,
  Fabrication, and Applications}}},\ \bibinfo {editor} {edited by\ \bibinfo
  {editor} {\bibfnamefont {S.}~\bibnamefont {Foner}}\ and\ \bibinfo {editor}
  {\bibfnamefont {B.~B.}\ \bibnamefont {Schwartz}}}\ (\bibinfo  {publisher}
  {Springer US},\ \bibinfo {address} {Boston, MA},\ \bibinfo {year} {1981})\
  pp.\ \bibinfo {pages} {275--388}\BibitemShut {NoStop}%
\bibitem [{\citenamefont {Davies}(2021)}]{Davies}%
  \BibitemOpen
  \bibfield  {author} {\bibinfo {author} {\bibfnamefont {A.}~\bibnamefont
  {Davies}},\ }\href
  {https://uspas.fnal.gov/materials/21onlineSBU/Background/Further%20reading%20-%20Cryogenic%20material%20properties.pdf}
  {\bibinfo {title} {Material properties data for heat transfer modeling in
  {Nb$_3$Sn} magnets}},\ \bibinfo {howpublished} {Review document for a course
  at the U.S. Particle Accelerator School (USPAS)} (\bibinfo {year}
  {2021})\BibitemShut {NoStop}%
\bibitem [{\citenamefont {Herzog}\ \emph {et~al.}(1982)\citenamefont {Herzog},
  \citenamefont {Khukhareva},\ and\ \citenamefont {Tsvineva}}]{Herzog}%
  \BibitemOpen
  \bibfield  {author} {\bibinfo {author} {\bibfnamefont {R.}~\bibnamefont
  {Herzog}}, \bibinfo {author} {\bibfnamefont {I.~S.}\ \bibnamefont
  {Khukhareva}},\ and\ \bibinfo {author} {\bibfnamefont {G.~P.}\ \bibnamefont
  {Tsvineva}},\ }\bibfield  {title} {\bibinfo {title} {Thermal conductivity of
  niobium–titanium alloys in the normal and superconducting states},\ }\href
  {https://doi.org/10.1063/10.0030772} {\bibfield  {journal} {\bibinfo
  {journal} {Soviet Journal Low Temperature Physics}\ }\textbf {\bibinfo
  {volume} {8}},\ \bibinfo {pages} {520} (\bibinfo {year} {1982})}\BibitemShut
  {NoStop}%
\bibitem [{\citenamefont {Mints}\ and\ \citenamefont {Brandt}(1996)}]{Mints1}%
  \BibitemOpen
  \bibfield  {author} {\bibinfo {author} {\bibfnamefont {R.~G.}\ \bibnamefont
  {Mints}}\ and\ \bibinfo {author} {\bibfnamefont {E.~H.}\ \bibnamefont
  {Brandt}},\ }\bibfield  {title} {\bibinfo {title} {Flux jumping in thin
  films},\ }\href {https://doi.org/10.1103/PhysRevB.54.12421} {\bibfield
  {journal} {\bibinfo  {journal} {Phys. Rev. B}\ }\textbf {\bibinfo {volume}
  {54}},\ \bibinfo {pages} {12421} (\bibinfo {year} {1996})}\BibitemShut
  {NoStop}%
\bibitem [{\citenamefont {Mints}\ and\ \citenamefont
  {Rakhmanov}(1981)}]{Mints2}%
  \BibitemOpen
  \bibfield  {author} {\bibinfo {author} {\bibfnamefont {R.~G.}\ \bibnamefont
  {Mints}}\ and\ \bibinfo {author} {\bibfnamefont {A.~L.}\ \bibnamefont
  {Rakhmanov}},\ }\bibfield  {title} {\bibinfo {title} {Critical state
  stability in {type-II} superconductors and superconducting-normal-metal
  composites},\ }\href {https://doi.org/10.1103/RevModPhys.53.551} {\bibfield
  {journal} {\bibinfo  {journal} {Rev. Mod. Phys.}\ }\textbf {\bibinfo {volume}
  {53}},\ \bibinfo {pages} {551} (\bibinfo {year} {1981})}\BibitemShut
  {NoStop}%
\bibitem [{\citenamefont {Swartz}\ and\ \citenamefont {Bean}(1968)}]{Swartz}%
  \BibitemOpen
  \bibfield  {author} {\bibinfo {author} {\bibfnamefont {P.~S.}\ \bibnamefont
  {Swartz}}\ and\ \bibinfo {author} {\bibfnamefont {C.~P.}\ \bibnamefont
  {Bean}},\ }\bibfield  {title} {\bibinfo {title} {A model for magnetic
  instabilities in hard superconductors: {T}he adiabatic critical state},\
  }\href {https://doi.org/10.1063/1.1655898} {\bibfield  {journal} {\bibinfo
  {journal} {J. Appl. Phys.}\ }\textbf {\bibinfo {volume} {39}},\ \bibinfo
  {pages} {4991} (\bibinfo {year} {1968})}\BibitemShut {NoStop}%
\bibitem [{\citenamefont {Gurevich}\ and\ \citenamefont
  {Mints}(1981)}]{Gurevich2}%
  \BibitemOpen
  \bibfield  {author} {\bibinfo {author} {\bibfnamefont {A.~V.}\ \bibnamefont
  {Gurevich}}\ and\ \bibinfo {author} {\bibfnamefont {R.~G.}\ \bibnamefont
  {Mints}},\ }\bibfield  {title} {\bibinfo {title} {Thermomagnetic effects,
  stability and oscillations in the critical state in hard superconductors},\
  }\href {https://doi.org/10.1088/0022-3727/14/6/021} {\bibfield  {journal}
  {\bibinfo  {journal} {Journal of Physics D: Applied Physics}\ }\textbf
  {\bibinfo {volume} {14}},\ \bibinfo {pages} {1129} (\bibinfo {year}
  {1981})}\BibitemShut {NoStop}%
\bibitem [{\citenamefont {Gurevich}\ and\ \citenamefont
  {Mints}(1987)}]{Gurevich}%
  \BibitemOpen
  \bibfield  {author} {\bibinfo {author} {\bibfnamefont {A.~V.}\ \bibnamefont
  {Gurevich}}\ and\ \bibinfo {author} {\bibfnamefont {R.~G.}\ \bibnamefont
  {Mints}},\ }\bibfield  {title} {\bibinfo {title} {Self-heating in normal
  metals and superconductors},\ }\href
  {https://doi.org/10.1103/RevModPhys.59.941} {\bibfield  {journal} {\bibinfo
  {journal} {Rev. Mod. Phys.}\ }\textbf {\bibinfo {volume} {59}},\ \bibinfo
  {pages} {941} (\bibinfo {year} {1987})}\BibitemShut {NoStop}%
\bibitem [{\citenamefont {Wipf}(1967)}]{Wipf}%
  \BibitemOpen
  \bibfield  {author} {\bibinfo {author} {\bibfnamefont {S.~L.}\ \bibnamefont
  {Wipf}},\ }\bibfield  {title} {\bibinfo {title} {Magnetic instabilities in
  type-{II} superconductors},\ }\href {https://doi.org/10.1103/PhysRev.161.404}
  {\bibfield  {journal} {\bibinfo  {journal} {Phys. Rev.}\ }\textbf {\bibinfo
  {volume} {161}},\ \bibinfo {pages} {404} (\bibinfo {year}
  {1967})}\BibitemShut {NoStop}%
\bibitem [{\citenamefont {Kramer}(1973)}]{Kramer}%
  \BibitemOpen
  \bibfield  {author} {\bibinfo {author} {\bibfnamefont {E.~J.}\ \bibnamefont
  {Kramer}},\ }\bibfield  {title} {\bibinfo {title} {Scaling laws for flux
  pinning in hard superconductors},\ }\href {https://doi.org/10.1063/1.1662353}
  {\bibfield  {journal} {\bibinfo  {journal} {J. Appl. Phys.}\ }\textbf
  {\bibinfo {volume} {44}},\ \bibinfo {pages} {1360} (\bibinfo {year}
  {1973})}\BibitemShut {NoStop}%
\bibitem [{\citenamefont {Nabia{\l}ek}\ \emph {et~al.}(2003)\citenamefont
  {Nabia{\l}ek}, \citenamefont {Chabanenko}, \citenamefont {Rusakov},
  \citenamefont {Vasiliev}, \citenamefont {Szymczak}, \citenamefont {Piechota},
  \citenamefont {Dabkowska}, \citenamefont {Dabkowski}, \citenamefont {Gaulin},
  \citenamefont {Niewczas},\ and\ \citenamefont {Mironov}}]{Nabialek}%
  \BibitemOpen
  \bibfield  {author} {\bibinfo {author} {\bibfnamefont {A.}~\bibnamefont
  {Nabia{\l}ek}}, \bibinfo {author} {\bibfnamefont {V.}~\bibnamefont
  {Chabanenko}}, \bibinfo {author} {\bibfnamefont {V.}~\bibnamefont {Rusakov}},
  \bibinfo {author} {\bibfnamefont {S.}~\bibnamefont {Vasiliev}}, \bibinfo
  {author} {\bibfnamefont {H.}~\bibnamefont {Szymczak}}, \bibinfo {author}
  {\bibfnamefont {S.}~\bibnamefont {Piechota}}, \bibinfo {author}
  {\bibfnamefont {H.}~\bibnamefont {Dabkowska}}, \bibinfo {author}
  {\bibfnamefont {A.}~\bibnamefont {Dabkowski}}, \bibinfo {author}
  {\bibfnamefont {B.~D.}\ \bibnamefont {Gaulin}}, \bibinfo {author}
  {\bibfnamefont {M.}~\bibnamefont {Niewczas}},\ and\ \bibinfo {author}
  {\bibfnamefont {O.}~\bibnamefont {Mironov}},\ }\bibfield  {title} {\bibinfo
  {title} {The peculiarities of magnetic flux dynamics at magnetothermal
  instability in textured {Bi$_2$Sr$_2$CaCu$_2$O$_{8+\delta}$}},\ }\href
  {https://doi.org/10.1023/A:1021847829436} {\bibfield  {journal} {\bibinfo
  {journal} {Journal of Low Temperature Physics}\ }\textbf {\bibinfo {volume}
  {130}},\ \bibinfo {pages} {425} (\bibinfo {year} {2003})}\BibitemShut
  {NoStop}%
\end{thebibliography}
\end{document}